\documentstyle[12pt,epsfig,axodraw]{article}

\advance\textheight by \topskip
\begin{document}
\tolerance=100000
\thispagestyle{empty}
\setcounter{page}{1}

\newcommand{\HPA}[1]{{\it Helv.\ Phys.\ Acta.\ }{\bf #1}}
\newcommand{\AP}[1]{{\it Ann.\ Phys.\ }{\bf #1}}
\newcommand{\be}{\begin{equation}}
\newcommand{\ee}{\end{equation}}
\newcommand{\br}{\begin{eqnarray}}
\newcommand{\er}{\end{eqnarray}}
\newcommand{\ba}{\begin{array}}
\newcommand{\ea}{\end{array}}
\newcommand{\bi}{\begin{itemize}}
\newcommand{\ei}{\end{itemize}}
\newcommand{\bn}{\begin{enumerate}}
\newcommand{\en}{\end{enumerate}}
\newcommand{\bc}{\begin{center}}
\newcommand{\ec}{\end{center}}
\newcommand{\ul}{\underline}
\newcommand{\ol}{\overline}
\def\l{\left\langle}
\def\r{\right\rangle}
\def\as{\alpha_{s}}
\def\ycut{y_{\mbox{\tiny cut}}}
\def\yij{y_{ij}}
\def\epem{\ifmmode{e^+ e^-} \else{$e^+ e^-$} \fi}
\newcommand{\eeww}{$e^+e^-\rightarrow W^+ W^-$}
\newcommand{\qqQQ}{$q_1\bar q_2 Q_3\bar Q_4$}
\newcommand{\eeqqQQ}{$e^+e^-\rightarrow q_1\bar q_2 Q_3\bar Q_4$}
\newcommand{\eewwqqqq}{$e^+e^-\rightarrow W^+ W^-\ar q\bar q Q\bar Q$}
\newcommand{\eeqqgg}{$e^+e^-\rightarrow q\bar q gg$}
\newcommand{\eeqloop}{$e^+e^-\rightarrow q\bar q gg$ via loop of quarks}
\newcommand{\eeqqqq}{$e^+e^-\rightarrow q\bar q Q\bar Q$}
\newcommand{\eewwjjjj}{$e^+e^-\rightarrow W^+ W^-\rightarrow 4~{\rm{jet}}$}
\newcommand{\eeqqggjjjj}{$e^+e^-\rightarrow q\bar 
q gg\rightarrow 4~{\rm{jet}}$}
\newcommand{\eeqloopjjjj}{$e^+e^-\rightarrow q\bar 
q gg\rightarrow 4~{\rm{jet}}$ via loop of quarks}
\newcommand{\eeqqqqjjjj}{$e^+e^-\rightarrow q\bar q Q\bar Q\rightarrow
4~{\rm{jet}}$}
\newcommand{\eejjjj}{$e^+e^-\rightarrow 4~{\rm{jet}}$}
\newcommand{\jjjj}{$4~{\rm{jet}}$}
\newcommand{\qqbar}{$q\bar q$}
\newcommand{\ww}{$W^+W^-$}
\newcommand{\ar}{\rightarrow}
\newcommand{\sm}{${\cal {SM}}$}
\newcommand{\Dir}{\kern -6.4pt\Big{/}}
\newcommand{\Dirin}{\kern -10.4pt\Big{/}\kern 4.4pt}
\newcommand{\DDir}{\kern -8.0pt\Big{/}}
\newcommand{\DGir}{\kern -6.0pt\Big{/}}
\newcommand{\wwqqqq}{$W^+ W^-\ar q\bar q Q\bar Q$}
\newcommand{\qqgg}{$q\bar q gg$}
\newcommand{\qloop}{$q\bar q gg$ via loop of quarks}
\newcommand{\qqqq}{$q\bar q Q\bar Q$}

\def\st{\sigma_{\mbox{\scriptsize t}}}
\def\Ord{\buildrel{\scriptscriptstyle <}\over{\scriptscriptstyle\sim}}
\def\OOrd{\buildrel{\scriptscriptstyle >}\over{\scriptscriptstyle\sim}}
\def\jhep #1 #2 #3 {{JHEP} {\bf#1} (#2) #3}
\def\plb #1 #2 #3 {{Phys.~Lett.} {\bf B#1} (#2) #3}
\def\npb #1 #2 #3 {{Nucl.~Phys.} {\bf B#1} (#2) #3}
\def\epjc #1 #2 #3 {{Eur.~Phys.~J.} {\bf C#1} (#2) #3}
\def\zpc #1 #2 #3 {{Z.~Phys.} {\bf C#1} (#2) #3}
\def\jpg #1 #2 #3 {{J.~Phys.} {\bf G#1} (#2) #3}
\def\prd #1 #2 #3 {{Phys.~Rev.} {\bf D#1} (#2) #3}
\def\prep #1 #2 #3 {{Phys.~Rep.} {\bf#1} (#2) #3}
\def\prl #1 #2 #3 {{Phys.~Rev.~Lett.} {\bf#1} (#2) #3}
\def\mpl #1 #2 #3 {{Mod.~Phys.~Lett.} {\bf#1} (#2) #3}
\def\rmp #1 #2 #3 {{Rev. Mod. Phys.} {\bf#1} (#2) #3}
\def\cpc #1 #2 #3 {{Comp. Phys. Commun.} {\bf#1} (#2) #3}
\def\sjnp #1 #2 #3 {{Sov. J. Nucl. Phys.} {\bf#1} (#2) #3}
\def\xx #1 #2 #3 {{\bf#1}, (#2) #3}
\def\hepph #1 {{\tt hep-ph/#1}}
\def\preprint{{preprint}}

\def\beq{\begin{equation}}
\def\beeq{\begin{eqnarray}}
\def\eeq{\end{equation}}
\def\eeeq{\end{eqnarray}}
\def\a0{\bar\alpha_0}
\def\thrust{\mbox{T}}
\def\Thrust{\mathrm{\tiny T}}
\def\ae{\alpha_{\mbox{\scriptsize eff}}}
\def\ap{\bar\alpha_p}
\def\as{\alpha_{\mathrm{S}}}
\def\aem{\alpha_{\mathrm{EM}}}
\def\b0{\beta_0}
\def\cN{{\cal N}}
\def\cd{\chi^2/\mbox{d.o.f.}}
\def\Ecm{E_{\mbox{\scriptsize cm}}}
\def\ee{e^+e^-}
\def\enap{\mbox{e}}
\def\eps{\epsilon}
\def\ex{{\mbox{\scriptsize exp}}}
\def\GeV{\mbox{\rm{GeV}}}
\def\half{{\textstyle {1\over2}}}
\def\jet{{\mbox{\scriptsize jet}}}
\def\kij{k^2_{\bot ij}}
\def\kp{k_\perp}
\def\kps{k_\perp^2}
\def\kt{k_\bot}
\def\lms{\Lambda^{(n_{\rm f}=4)}_{\overline{\mathrm{MS}}}}
\def\mI{\mu_{\mathrm{I}}}
\def\mR{\mu_{\mathrm{R}}}
\def\MSbar{\overline{\mathrm{MS}}}
\def\mx{{\mbox{\scriptsize max}}}
\def\NP{{\mathrm{NP}}}
\def\pd{\partial}
\def\pt{{\mbox{\scriptsize pert}}}
\def\pw{{\mbox{\scriptsize pow}}}
\def\so{{\mbox{\scriptsize soft}}}
\def\st{\sigma_{\mbox{\scriptsize tot}}}
\def\ycut{y_{\mathrm{cut}}}
\def\slashchar#1{\setbox0=\hbox{$#1$}           
     \dimen0=\wd0                                 
     \setbox1=\hbox{/} \dimen1=\wd1               
     \ifdim\dimen0>\dimen1                        
        \rlap{\hbox to \dimen0{\hfil/\hfil}}      
        #1                                        
     \else                                        
        \rlap{\hbox to \dimen1{\hfil$#1$\hfil}}   
        /                                         
     \fi}                                         %
\def\etmiss{\slashchar{E}^T}
\def\Meff{M_{\rm eff}}
\def\Ord{\lsim}
\def\OOrd{\gsim}
\def\tq{\tilde q}
\def\tchi{\tilde\chi}
\def\lsp{\tilde\chi_1^0}

\def\gam{\gamma}
\def\ph{\gamma}
\def\be{\begin{equation}}
\def\ee{\end{equation}}
\def\bea{\begin{eqnarray}}
\def\eea{\end{eqnarray}}
\def\lsim{\:\raisebox{-0.5ex}{$\stackrel{\textstyle<}{\sim}$}\:}
\def\gsim{\:\raisebox{-0.5ex}{$\stackrel{\textstyle>}{\sim}$}\:}

\def\ino{\mathaccent"7E} \def\gluino{\ino{g}} \def\mgluino{m_{\gluino}}
\def\sqk{\ino{q}} \def\sup{\ino{u}} \def\sdn{\ino{d}}
\def\chargino{\ino{\omega}} \def\neutralino{\ino{\chi}}
\def\cab{\ensuremath{C_{\alpha\beta}}} \def\proj{\ensuremath{\mathcal P}}
\def\dab{\delta_{\alpha\beta}}
\def\zz{s-M_Z^2+iM_Z\Gamma_Z} \def\zw{s-M_W^2+iM_W\Gamma_W}
\def\prop{\ensuremath{\mathcal G}} \def\ckm{\ensuremath{V_{\rm CKM}^2}}
\def\aem{\alpha_{\rm EM}} \def\stw{s_{2W}} \def\sttw{s_{2W}^2}
\def\nc{N_C} \def\cf{C_F} \def\ca{C_A}
\def\qcd{\textsc{Qcd}} \def\susy{supersymmetric} \def\mssm{\textsc{Mssm}}
\def\slash{/\kern -5pt} \def\stick{\rule[-0.2cm]{0cm}{0.6cm}}
\def\h{\hspace*{-0.3cm}}

\def\ims #1 {\ensuremath{M^2_{[#1]}}}
\def\tw{\tilde \chi^\pm}
\def\tz{\tilde \chi^0}
\def\tf{\tilde f}
\def\tl{\tilde l}
\def\ppb{p\bar{p}}
\def\gl{\tilde{g}}
\def\sq{\tilde{q}}
\def\sqb{{\tilde{q}}^*}
\def\qb{\bar{q}}
\def\sqL{\tilde{q}_{_L}}
\def\sqR{\tilde{q}_{_R}}
\def\ms{m_{\tilde q}}
\def\mg{m_{\tilde g}}
\def\Gs{\Gamma_{\tilde q}}
\def\Gg{\Gamma_{\tilde g}}
\def\md{m_{-}}
\def\eps{\varepsilon}
\def\Ce{C_\eps}
\def\dnq{\frac{d^nq}{(2\pi)^n}}
\def\DR{$\overline{DR}$\,\,}
\def\MS{$\overline{MS}$\,\,}
\def\DRm{\overline{DR}}
\def\MSm{\overline{MS}}
\def\ghat{\hat{g}_s}
\def\shat{\hat{s}}
\def\sihat{\hat{\sigma}}
\def\Li{\text{Li}_2}
\def\bs{\beta_{\sq}}
\def\xs{x_{\sq}}
\def\xsa{x_{1\sq}}
\def\xsb{x_{2\sq}}
\def\bg{\beta_{\gl}}
\def\xg{x_{\gl}}
\def\xga{x_{1\gl}}
\def\xgb{x_{2\gl}}
\def\lsp{\tilde{\chi}_1^0}

\def\gluino{\mathaccent"7E g}
\def\mgluino{m_{\gluino}}
\def\squark{\mathaccent"7E q}
\def\msquark{m_{\mathaccent"7E q}}
\def\M{ \overline{|\mathcal{M}|^2} }
\def\utm{ut-M_a^2M_b^2}
\def\MiLR{M_{i_{L,R}}}
\def\MiRL{M_{i_{R,L}}}
\def\MjLR{M_{j_{L,R}}}
\def\MjRL{M_{j_{R,L}}}
\def\tiLR{t_{i_{L,R}}}
\def\tiRL{t_{i_{R,L}}}
\def\tjLR{t_{j_{L,R}}}
\def\tjRL{t_{j_{R,L}}}
\def\tg{t_{\gluino}}
\def\uiLR{u_{i_{L,R}}}
\def\uiRL{u_{i_{R,L}}}
\def\ujLR{u_{j_{L,R}}}
\def\ujRL{u_{j_{R,L}}}
\def\ug{u_{\gluino}}
\def\utot{u \leftrightarrow t}
\def\ar{\to}
\def\sqk{\mathaccent"7E q}
\def\sup{\mathaccent"7E u}
\def\sdn{\mathaccent"7E d}
\def\chargino{\mathaccent"7E \chi}
\def\neutralino{\mathaccent"7E \chi}
\def\slepton{\mathaccent"7E l}
\def\M{ \overline{|\mathcal{M}|^2} }
\def\cab{\ensuremath{C_{\alpha\beta}}}
\def\ckm{\ensuremath{V_{\rm CKM}^2}}
\def\zz{s-M_Z^2+iM_Z\Gamma_Z}
\def\zw{s-M_W^2+iM_W\Gamma_W}
\def\s22w{s_{2W}^2}

\newcommand{\cpmtwo}    {\mbox{$ {\chi}^{\pm}_{2}                    $}}
\newcommand{\cpmone}    {\mbox{$ {\chi}^{\pm}_{1}                    $}}

\begin{flushleft}
{SHEP-03-37}\\
{DFTT 06/2004}
\end{flushleft}
\begin{flushright}
\vskip-1.55truecm
{LC-TH-2004-005\\
March 2004}
\end{flushright}
\vskip0.1cm\noindent
\begin{center}
{\Large {\bf One-loop weak corrections \\[0.25cm]
to three-jet observables at the 
$Z$ pole\footnote{Work supported in 
part by the U.K.\ Particle Physics and
Astronomy Research Council (PPARC),
by the European Union (EU) under contract HPRN-CT-2000-00149 and by the 
Italian Ministero dell'Istruzione, dell'Universit\`a e della Ricerca
(MIUR) under contract 2001023713\_006.}}}
\\[1.5cm]
{\large Ezio Maina}\\[0.15 cm]
{\it Dipartimento di Fisica Teorica -- Universit\`a di Torino}\\
{\it and} \\
{\it Istituto Nazionale di Fisica Nucleare -- Sezione di Torino}\\
{\it Via Pietro Giuria 1, 10125 Torino, Italy}
\\[0.5cm]
{\large Stefano Moretti\footnote{Talk given at the ECFA Study of Physics and Detectors for a Linear Collider, Montpellier,
France, 13-16 November 2003.} and Douglas A. Ross}\\[0.15 cm]
{\it School of Physics and Astronomy, University of Southampton}\\
{\it Highfield, Southampton SO17 1BJ, UK}\\[0.25cm]
\end{center}

\begin{abstract}
{\small
\noindent
We briefly illustrate the impact of the genuinely weak one-loop 
virtual terms of the ${\cal O}(\alpha_{\rm S}\alpha_{\rm EW}^3)$ 
factorisable corrections
to the three-jet cross section at $\sqrt s=M_Z$. Their
importance for the measurement of $\alpha_{\rm S}$ 
at GigaZ luminosities is emphasised.}
\end{abstract}

\section{Three-jets events at leptonic colliders}
\label{Sec:ee}
 
In the case of $e^+e^-$ annihilations, the most important QCD quantity to be 
extracted from multi-jet events is $\alpha_{\mathrm{S}}$.
The confrontation of the measured value of the strong coupling
constant with that predicted by the theory through the 
renormalisation group evolution is an important test of the Standard Model
(SM) or else an indication of new physics, when its typical mass scale is 
larger than the collider energy, 
but which can manifest itself through virtual effects. Not only jet rates,
but also
jet shape observables, which offer a handle on non-perturbative
QCD effects via large power corrections, would be affected.

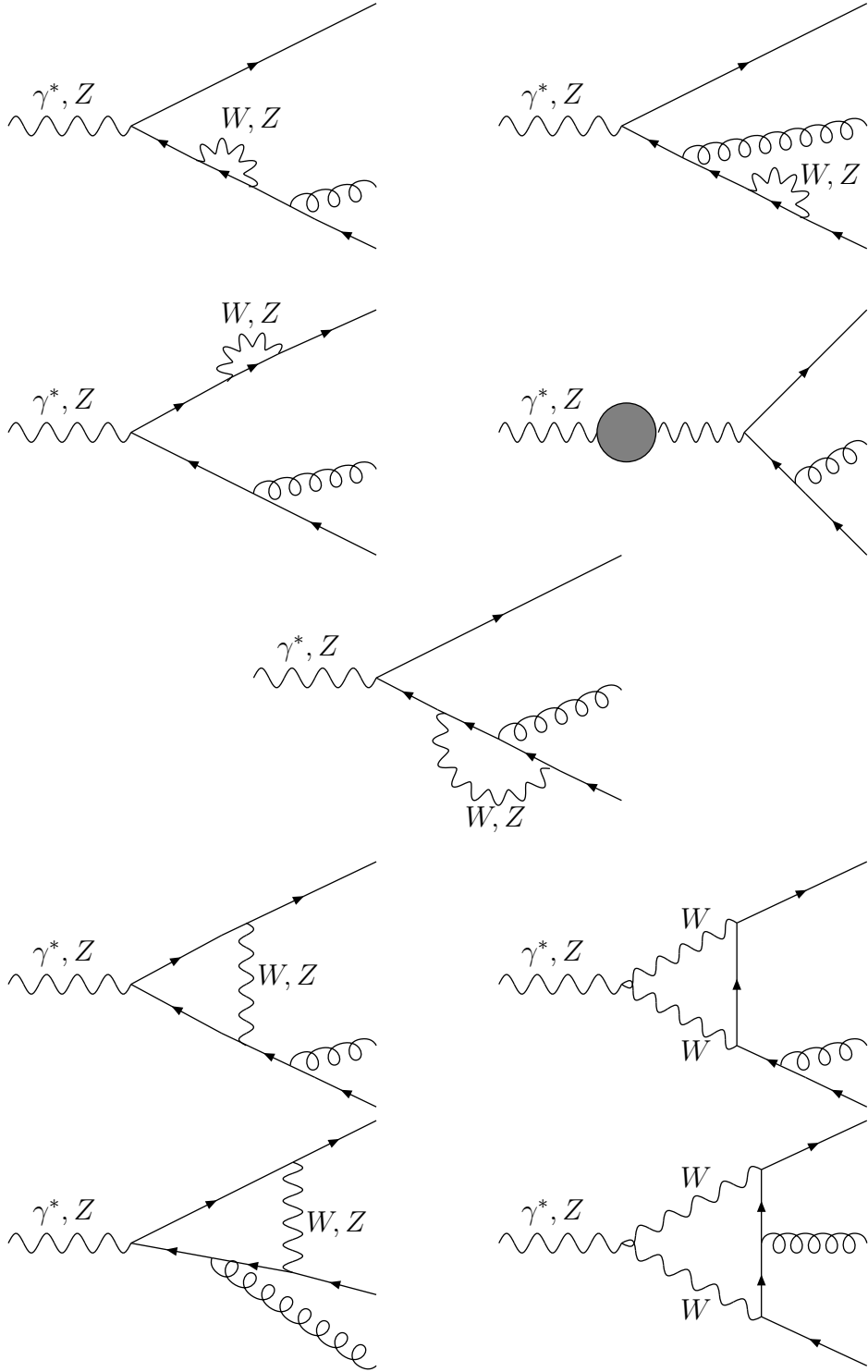
\begin{figure}[!h]
\begin{center}
\begin{picture}(365,225)
\SetScale{1.0}
\Photon(0,175)(50,175){4}{4}
\ArrowLine(150,125)(125,137) \Line(125,137)(100,150)
\ArrowLine(100,150)(75,162) \ArrowLine(75,162)(50,175)
\ArrowLine(50,175)(150,225) \Gluon(115,142)(150,150){4}{3}
\PhotonArc(87,155)(12,337,150){3}{5} \put(87,175){$W,Z$}
\put(10,185){$\gamma^*,Z$}

\Photon(200,175)(250,175){4}{4}
\ArrowLine(350,125)(325,137) \ArrowLine(325,137)(300,150)
\ArrowLine(300,150)(275,162) \ArrowLine(275,162)(250,175)
\ArrowLine(250,175)(350,225) \Gluon(275,162)(350,175){4}{8}
\PhotonArc(312,143)(12,334,150){3}{5} \put(324,152){$W,Z$}
\put(210,185){$\gamma^*,Z$}

\Photon(0,50)(50,50){4}{4}
\ArrowLine(150,0)(100,25) \ArrowLine(100,25)(50,50)
\ArrowLine(50,50)(90,72) \ArrowLine(90,72)(110,82)
\ArrowLine(110,82)(150,100) \Gluon(100,25)(150,35){4}{5}
\PhotonArc(100,76)(12,32,206){3}{5} \put(87,95){$W,Z$}
\put(10,60){$\gamma^*,Z$}

\Photon(200,50)(240,50){4}{4} \Photon(265,50)(300,50){4}{4}
\ArrowLine(350,0)(325,25) \ArrowLine(325,25)(300,50)
\ArrowLine(300,50)(350,100) \Gluon(321,29)(350,45){4}{3}
\GCirc(252,50){12}{.5}
\put(210,60){$\gamma^*,Z$}
\end{picture}

\begin{picture}(365,225)

\Photon(100,175)(150,175){4}{4}
\ArrowLine(250,125)(225,137) \ArrowLine(225,137)(200,150)
\ArrowLine(200,150)(175,162) \ArrowLine(175,162)(150,175)
\ArrowLine(150,175)(250,225) \Gluon(200,150)(250,170){4}{5}
\PhotonArc(200,150)(24,154,330){3}{8} \put(187,115){$W,Z$}
\put(110,185){$\gamma^*,Z$}

\Photon(0,50)(50,50){4}{4}
\ArrowLine(150,0)(125,12) \ArrowLine(125,12)(87,30)
 \ArrowLine(87,30)(50,50)
\ArrowLine(50,50)(87,70) \ArrowLine(87,70)(150,100)
 \Gluon(115,17)(150,25){4}{3}
\Photon(97,25)(97,75){3}{5} \put(102,50){$W,Z$}
\put(10,60){$\gamma^*,Z$}

\Photon(200,50)(250,50){4}{4}
\ArrowLine(350,0)(325,12) \ArrowLine(325,12)(297,25)
 \Photon(297,25)(250,50){3}{5}
\Photon(250,50)(297,75){-3}{5} \ArrowLine(297,75)(350,100)
 \Gluon(315,17)(350,25){4}{3}
\ArrowLine(297,25)(297,75) \put(275,73){$W$} \put(275,19){$W$}
\put(210,60){$\gamma^*,Z$}

\end{picture}

\begin{picture}(365,105)

\Photon(0,50)(50,50){4}{4}
\put(10,60){$\gamma^*,Z$}
\ArrowLine(150,29)(116,38) \ArrowLine(116,38)(83,44) \ArrowLine(83,44)(50,50)
\ArrowLine(50,50)(116,83) \ArrowLine(116,83)(150,100)
\Gluon(83,44)(150,0){-4}{8} \Photon(116,38)(116,83){4}{5}
\put(122,55){$W,Z$}

\Photon(200,50)(250,50){4}{4}
\ArrowLine(350,0)(307,20) 
 \Photon(307,20)(250,50){3}{5}
\Photon(250,50)(307,80){-3}{5} \ArrowLine(307,80)(350,100)
 \Gluon(307,50)(350,50){4}{5}
\ArrowLine(307,20)(307,50) \ArrowLine(307,50)(307,80)
 \put(275,73){$W$} \put(275,19){$W$}
\put(210,60){$\gamma^*,Z$}

\end{picture}
\end{center}
\caption{\small Graphs describing $\gamma^*,Z\to \bar qqg$ in presence
of one-loop weak corrections.}
\label{fig:graphs}
\end{figure}

We report here on the calculation of the
one-loop weak-interaction corrections
to three-jet observables in electron-positron annihilations
through the order ${\cal O}(\alpha_{\rm S}\alpha_{\rm EW}^3)$
generated via the interference of the graphs
in Fig.~\ref{fig:graphs} with the tree-level ones for 
$\gamma^*,Z\to \bar q qg$
final states, where the external $\gamma^*,Z$ current is intended to
connect to an incoming $e^+e^-$ fermion line. In the figure, the shaded blob
represents all the contributions to the gauge boson 
self-energy and is dependent on the Higgs mass (we have set
$M_H=115$ GeV for the latter). We neglect, however, loops involving the
Higgs boson coupling to the fermion lines. The graphs in which the 
exchanged gauge boson is a $W$-boson
are accompanied by those in which the latter is replaced
by its corresponding Goldstone boson. There is then a similar set of diagrams
in which the direction of the fermion line is reversed, with the exception
of the last graph, as here reversal does {not} lead to a distinct topology.
Corrections along the $e^+e^-$ lines are also included in our analysis. 
In short, we compute the factorisable effects only,
i.e., corrections to the initial and 
final states only.
Whereas these should be sufficient to describe adequately
the phenomenology of three-jet events at $\sqrt s=M_Z$, at energy
scales much larger than the $Z$-boson mass
one expects comparable effects due to the non-factorisable corrections,
in which weak gauge-bosons connect via one-loop diagrams electrons and 
positrons to quarks and antiquarks, whose computation is currently 
in progress.

We will show that such factorisable corrections are of a few percent at 
$\sqrt s=M_Z$. Hence, while their impact is not dramatic in the context
of LEP1 and SLC, where the total error on the measured value of
$\alpha_{\mathrm{S}}$ is larger, at a future 
Linear Collider (LC) \cite{LCs}, running at the $Z$ mass peak (e.g., GigaZ),
they ought to be taken into account in the experimental fits,
as here the uncertainty on the value of the strong coupling
constant is expected to be at the $0.1\%$ level or even smaller \cite{Winter}.
The calculation has been performed using helicity amplitudes so that
it can be applied to the case of polarised beams, option that is
one of the strengths of the LC projects. Besides, another
aspect that should be recalled is that weak corrections naturally
introduce parity-violating effects in jet observables, detectable through
asymmetries in the cross-section, which are often regarded as an indication
of physics beyond the SM. Comparison of theoretical predictions 
involving parity-violation with future LC experimental data 
is regarded as another powerful tool for confirming or 
disproving  the existence of some beyond the SM scenarios, such as those 
involving right-handed weak currents and/or new massive gauge bosons.

\section{Calculation}
\label{Sec:Calculation}

For the techniques used in the calculation of the one-loop diagrams in 
Fig.~\ref{fig:graphs}, as well as for the relevant SM input parameters,
we refer the reader to ~\cite{Maina:2002wz}. 

The three-parton cross section for 
$e^+e^-\to \bar q q g$ can be written in terms
of nine form-factors  \cite{BDS,KS}. The latter
generate the doubly differential cross-section for 
three-jet production in terms of some event shape variable, $S$, 
which is in turn related to  the energy
fractions $x_1, \, x_2$ of the antiquark and quark, respectively
($x_i=\frac{2E_i}{\sqrt s}$), 
by some function, $s$, i.e., $S = s(x_1,x_2) $,
and of the polar and azimuthal angles, $\alpha, \, \beta$, 
between, e.g., the incoming electron beam and the antiquark jet: 
\br\label{FFs}
 \frac{d^3\sigma}{dS \, d\cos\alpha \, d\beta} & =  &\int dx_1 dx_2
 ~\delta \left(S-s(x_1,x_2) \right)    
\left[ (2-\sin^2\alpha) \, F_1  +(1-3\cos^2\alpha) 
\, F_2
\right. \nonumber \\ 
 & &  \hspace*{1cm}  + \, \lambda_e \,
 \cos\alpha \, F_3  + \sin 2\alpha \cos\beta \, F_4 
+ \sin^2\alpha \cos 2\beta \, F_5 + \lambda_e \,
 \sin\alpha \cos\beta \, F_6 \nonumber \\
& & \hspace*{1 cm} + \, \sin 2 \alpha \sin \beta \, F_7 + \sin^2 \alpha \sin 2 \beta \,
 F_8 
\left. + \, \lambda_e \, \sin\alpha \sin\beta \, F_9 \right] . \label{eq35}
 \er  
The last three terms ($F_7, ... F_9$) arise for the first time at the 
one-loop 
level, since they are proportional to the imaginary parts of the 
helicity-matrix\footnote{For $F_9$, this is strictly true only for massless
quarks as, for $m_q\ne0$, Ref.~\cite{BDS} has shown that this form-factor
becomes non-zero also in pure QCD.}.
$F_3, \ F_6$ and $F_7$ vanish in the parity-conserving limit and 
can therefore be used
as probes of weak interaction contributions to three-jet production.
Moreover, $F_3$ and $F_6$ would be exactly 
 zero at tree level if the leading order 
process were only mediated by virtual photons. Finally, notice
 that upon integrating over the antiquark angle 
relative to the electron
beam, only the form-factor $F_1$ survives. (The expression of the
nine form-factors in terms of the helicity amplitudes can be
found in Ref.~\cite{Maina:2002wz}.)

In general, it is not possible to distinguish between quark, antiquark
and gluon jets, although the above expression can easily be adapted such that 
the angles $\alpha, \beta$ refer to the leading jet. However, (anti)quark
jets {can} be recognised when they originate from primary 
$b$-(anti)quarks, thanks to rather efficient flavour tagging techniques
(such as  $\mu$-vertex devices). We will therefore consider 
the numerical results for such a case separately.

\section{Numerical results}
\label{Sec:Results}

The processes considered here are the following: 
\begin{equation}\label{procj}
e^+e^-\to \gamma^*,Z\to \bar qqg\quad{\mathrm{(all~flavours)}},
\end{equation}
when no assumption is made on the flavour content of the final state,
so that a summation will be performed over $q=d,u,c,s,b$-quarks, and 
\begin{equation}\label{procb}
e^+e^-\to \gamma^*,Z\to \bar bbg,
\end{equation}
limited to the case of bottom quarks only in the final state. 
All quarks
in the final state of (\ref{procj})--(\ref{procb}) are taken as 
massless\footnote{Mass effects in $e^+e^-\to \gamma^*,Z^{(*)}\to \bar bbg$
have been studied in \cite{BMM} and \cite{bbgNLO}.}.
In contrast, the top quark entering the loops in both reactions has
been assumed to be massive.  
We can systematically neglect higher order effects from Electro-Magnetic
(EM) radiation without incurring into gauge-invariance violations. We
do not include  beamstrahlung corrections either. 

It is common in the specialised literature to define the $n$-jet 
fraction $R_n(y)$ as
\beq
\label{fn}
R_n(y)=\frac{\sigma_n(y)}{\sigma_0},
\eeq
where $y$  is a suitable
variable quantifying the space-time separation among hadronic objects
and with $\sigma_{0}$ identifying the (energy-dependent) 
Born cross-section for $e^+e^-\to \bar qq$.
For the choice $\mu=\sqrt s$ of the renormalisation scale, 
one can conveniently write the three-jet fraction in the following form:
\beq
\label{f3}
R_3(y) =     \left( \frac{\as}{2\pi} \right)    A(y)
           + \left( \frac{\as}{2\pi} \right)^2  B(y) + ... ,
\eeq
where the coupling constant $\as$ and the functions $A(y)$ and $B(y)$ 
are defined in  the $\overline{\mbox{MS}}$ scheme. An experimental fit
of the $R_n(y)$ jet fractions to the corresponding 
theoretical prediction is a powerful
way of determining $\as$ from multi-jet rates. Leading-order (LO) terms
enter $A(y)$ while next-to-leading order QCD effects (hereafter, labelled as 
NLO-QCD) are embedded in $B(y)$, see \cite{ERT}, with contribution from
both (genuine) three- and (unresolved) four-parton final states.
The weak corrections of interest (hereafter,
labelled as NLO-W) only contribute to the former. Hence,
in order to account for these, it suffices to make the replacement
\beq
\label{f3EW}
A(y)\to A(y)+A_{\mathrm{W}}(y)
\eeq
in eq.~(\ref{f3}). The jet multiplicity of the hadronic final state 
is determined through the implementation of
so-called `jet clustering algorithms' (see \cite{schemes}
for a description of their properties). Among the latter
we use the JADE (J) \cite{jade}, Durham (D) \cite{durham}, Cambridge (C)
\cite{cambridge} and Geneva (G) \cite{BKSS} ones.

\begin{figure}
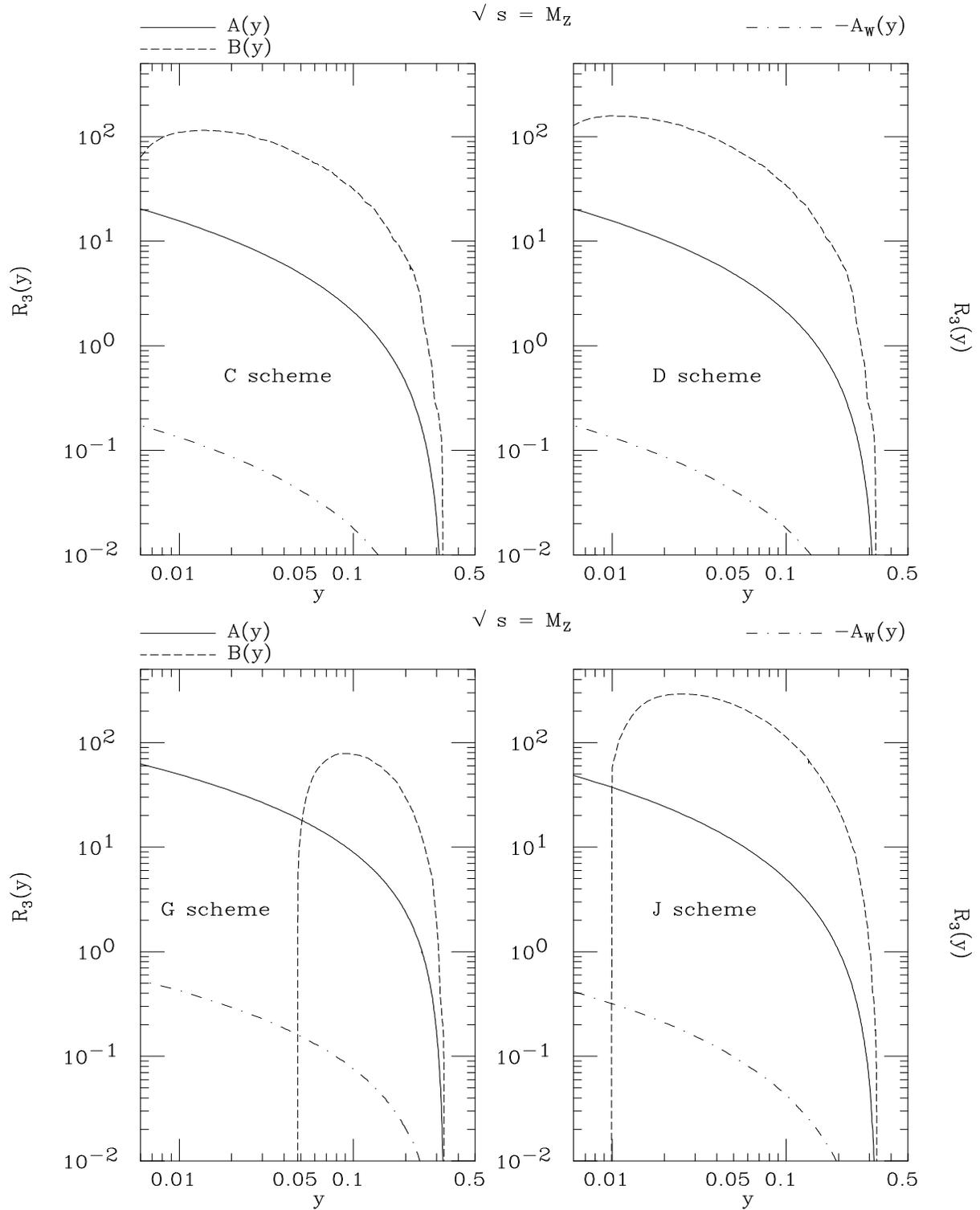

\begin{center}
\epsfig{file=yCD_LEP1_new.ps,height=160mm, width=100mm, angle=90}
\epsfig{file=yGJ_LEP1_new.ps,height=160mm, width=100mm, angle=90}
\end{center}
\vskip -0.5cm
\caption{\small The $A(y)$, $-A_{\mathrm{W}}$ and $B(y)$ coefficient functions
of eqs.~(\ref{f3})--(\ref{f3EW}) for the Cambridge, Durham, Geneva and
Jade jet clustering algorithms, at $\sqrt s=M_Z$. (Notice that
the $A_{\mathrm{W}}$ term
has been plotted with opposite sign
for better presentation.)}
\label{fig:y_LEP1}
\end{figure}

Fig.~\ref{fig:y_LEP1} displays\footnote{Hereafter, NLO QCD
results are obtained from EERAD \cite{EERAD}.} the $A(y)$, $-A_{\rm{W}}(y)$ 
and $B(y)$ coefficients 
entering eqs.~(\ref{f3})--(\ref{f3EW}), as a function of 
$y$ (a jet separation parameter \cite{schemes}) for the above four 
jet algorithms at $\sqrt s=M_Z$. (Notice 
that $A(y)$ and $A_{\rm{W}}(y)$ for the
C scheme are identical to those for the D one\footnote{The Cambridge 
algorithm in 
fact only modifies the clustering procedure of the Durham jet finder and the
two implementations coincide for $n\le 3$ parton final states,
as they use the same separation variable \cite{schemes}.}.)
 A comparison
between $A(y)$ and $A_{\rm{W}}(y)$ reveals that the NLO-W corrections are
negative and remain
at the percent level, i.e., of order $\frac{\aem}{2\pi s_W^2}$.
They give rise to corrections to $\sigma_3(y)$ of --1\%, and
thus are generally much smaller than the NLO-QCD ones. In this context, no
systematic difference is seen with respect to the choice of jet clustering
algorithm, over the typical range of application of the latter at $\sqrt
s=M_Z$ (say, 
$\ycut\gsim0.005$ for D, C and $\ycut\gsim0.01$ for G, J).

\begin{figure}
\begin{center}
\hskip -0.75cm\epsfig{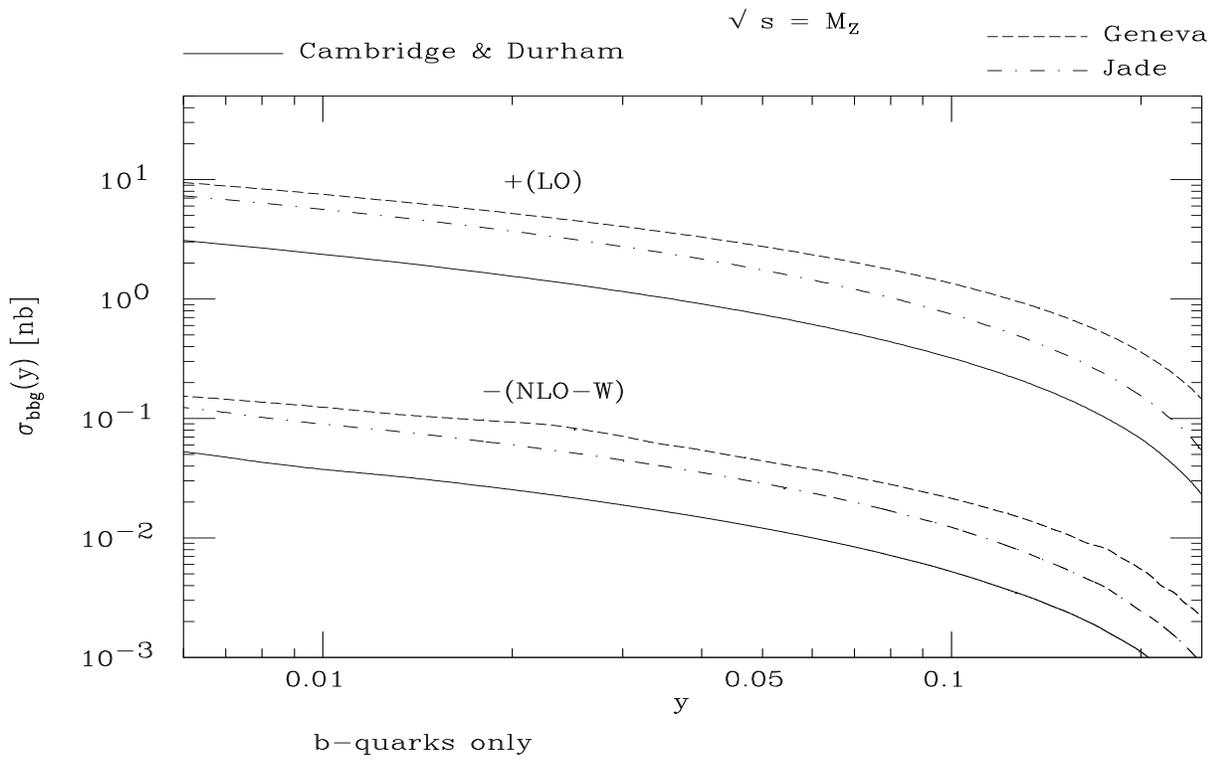}
\end{center}
\vskip -0.5cm
\caption{\small The total cross section for process (\ref{procb}) at LO
and NLO-W for the Cambridge, Durham, Geneva and
Jade jet clustering algorithms, at $\sqrt s=M_Z$. (Notice that the
NLO-W results have been plotted with opposite sign
for better presentation.)}
\label{fig:y_LEP1_b}
\end{figure}

As already mentioned, it should
now be recalled that jets originating from $b$-quarks can efficiently be
distinguished from light-quark jets. 
Besides, the $b$-quark component of the full three-jet sample is the
only one sensitive to $t$-quark loops, hence one may expect somewhat
different effects from weak corrections to process (\ref{procb})
than to (\ref{procj}) (the residual dependence on the $Z \bar q q$
couplings is also different). This is confirmed by 
Fig.~\ref{fig:y_LEP1_b}, where we present the total cross section at $\sqrt
s=M_Z$ for $e^+e^-\to\gamma^*,Z\to\bar bbg$ as obtained at LO and NLO-W, for
our usual choice of jet clustering algorithms and separations. A close
inspection of the plots reveals that NLO-W effects can reach the 
$-2.0\%$ level or so.

In view of these percent effects being well above the error estimate
expected at a future high-luminosity LC running at the $Z$ pole,
it is then worthwhile to further consider the effects
of NLO-W corrections to some other `infrared-safe' jet observables typically
used in the determination
of $\as$, the so-called `shape variables'. 
A representative quantity in this respect is the Thrust (T)
distribution \cite{thrust}. This is defined as the sum of 
the longitudinal momenta relative to the (Thrust) axis $n_{\rm T}$ chosen
to maximise this sum, i.e.:
\begin{equation}\label{thrust}
\thrust = {\rm max} \frac{\sum_i |\vec{p_i}\cdot\vec{n_{\mathrm{T}}}|}
                         {\sum_i |\vec{p_i}|} ,
\end{equation} 
where $i$ runs over all final state objects.
This quantity is identically one at Born level, getting
the first non-trivial
contribution through ${\cal O}(\as\alpha_{\rm EW}^2)$ from events of the
type (\ref{procj})--(\ref{procb}). Also notice that any other higher 
order contribution will affect this observable. Through 
${\cal O}(\as^2\alpha_{\rm EW}^2)$,
for the choice $\mu =\sqrt s$ of the renormalisation scale, 
the T distribution can be parametrised in the following form:
\begin{equation}\label{T}
(1-{\rm{T}})\frac{d\sigma}{d\thrust}\frac{1}{\sigma_0} = 
\left(\frac{\as}{2\pi}\right)   A^{\Thrust}(\thrust)+
\left(\frac{\as}{2\pi}\right)^2 B^{\Thrust}(\thrust).
\end{equation} 
Again, the replacement 
\begin{equation}\label{TW}
A^{\Thrust}(\thrust)\to A^{\Thrust}(\thrust)+A^{\Thrust}_{\rm{W}}(\thrust) 
\end{equation} 
accounts for the inclusion of the NLO-W contributions.

\begin{figure}
\begin{center}
\epsfig{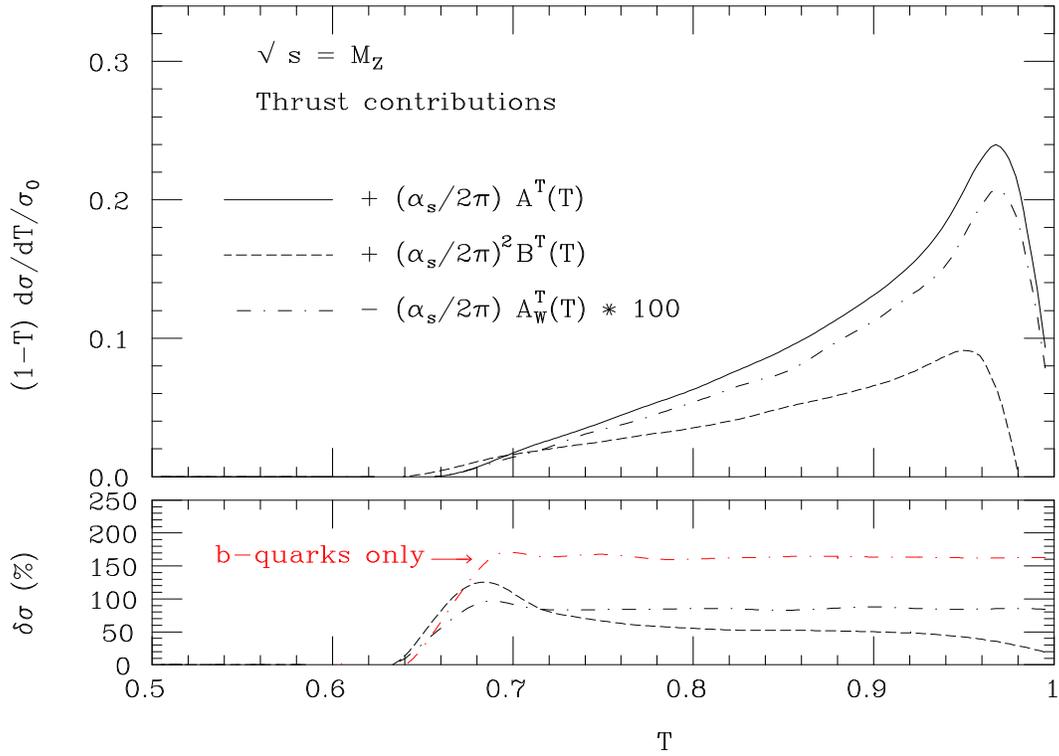}
\end{center}
\vskip -0.5cm
\caption{\small The LO, NLO-QCD and NLO-W  
contributions to the coefficient functions entering the
integrated Thrust distribution, see eq.~(\ref{T}), for
process (\ref{procj}) (top) and the relative 
size of the two NLO corrections (bottom), at $\sqrt s=M_Z$.
The correction for the case of $b$-quarks only is also
presented, relative to the LO results for process (\ref{procb}).
 (Notice that the
$A_{\mathrm{W}}$ terms
have been plotted with opposite sign and
multiplied by hundred 
for better presentation.)}
\label{fig:thrust}
\end{figure}

We plot the terms $\left(\frac{\as}{2\pi}\right)A^{\Thrust}(\thrust)$,
$\left(\frac{\as}{2\pi}\right)A^{\Thrust}_{\rm{W}}(\thrust)$ and 
$\left(\frac{\as}{2\pi}\right)^2B^{\Thrust}(\thrust)$ in Fig.~\ref{fig:thrust},
always at  $\sqrt s=M_Z$,
alongside the relative rates of the NLO-QCD and NLO-W terms 
with respect to the LO contribution. Here, it can be seen that
the NLO-W effects can reach the level of $-1\%$ or so and that they are
fairly constant for $0.7\lsim {\rm{T}}\lsim 1$. For the case of $b$-quarks
only, similarly to what seen already for the inclusive
rates, the NLO-W corrections are larger, as they can reach the $-1.6\%$ level.

\begin{figure}
\begin{center}
\centerline{}
\begin{minipage}[b]{.33333\linewidth}
\centerline{$-F_1$}
\vspace{-0.75truecm}\centering\epsfig{file=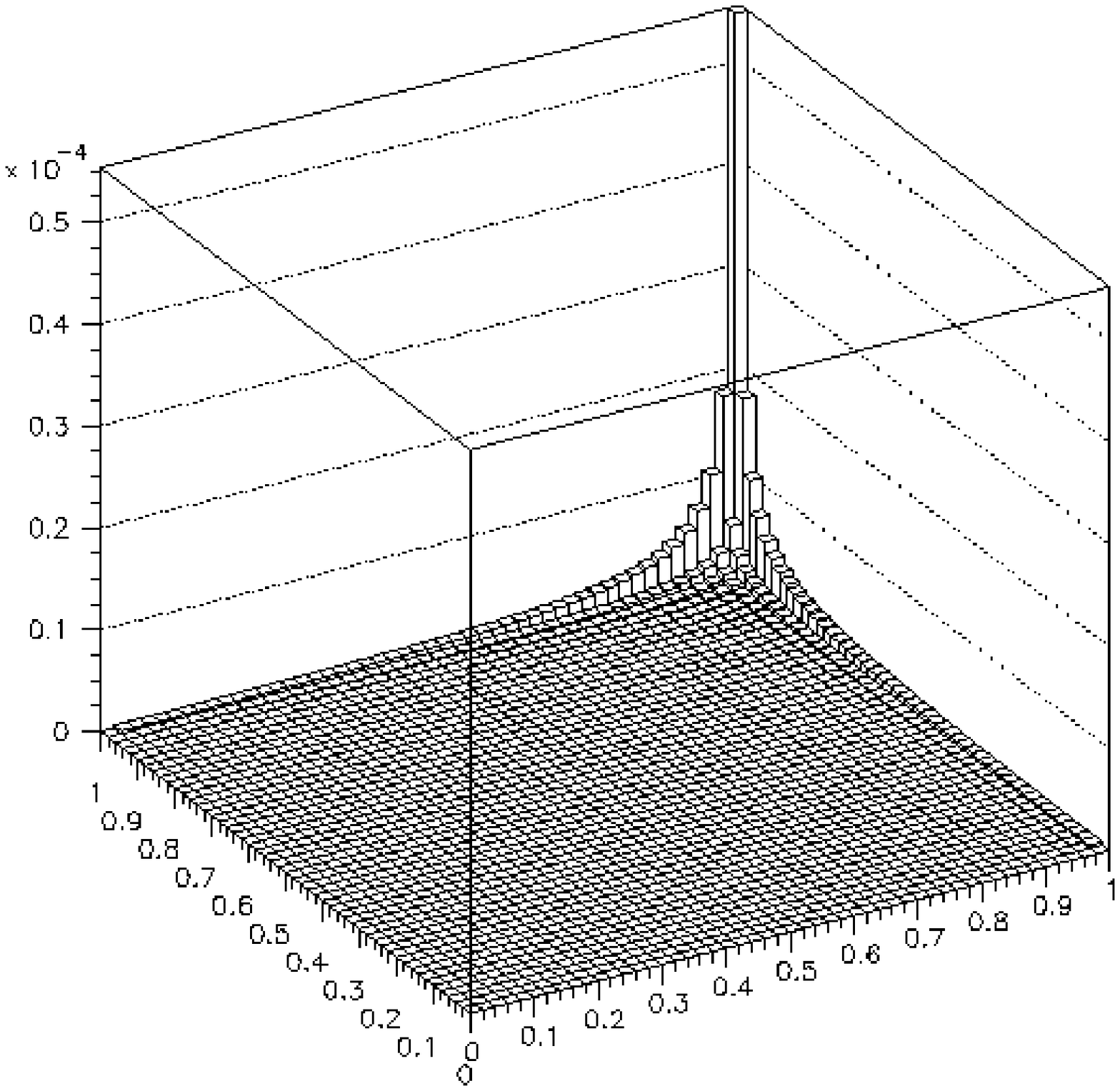,angle=0,height=8cm,width=\linewidth}
\vskip-2.0cm
\centerline{
$x_2$
\qquad\qquad\qquad\qquad
$x_1$}
\end{minipage}\hfil
\begin{minipage}[b]{.33333\linewidth}
\centerline{$-F_2$}
\vspace{-0.75truecm}\centering\epsfig{file=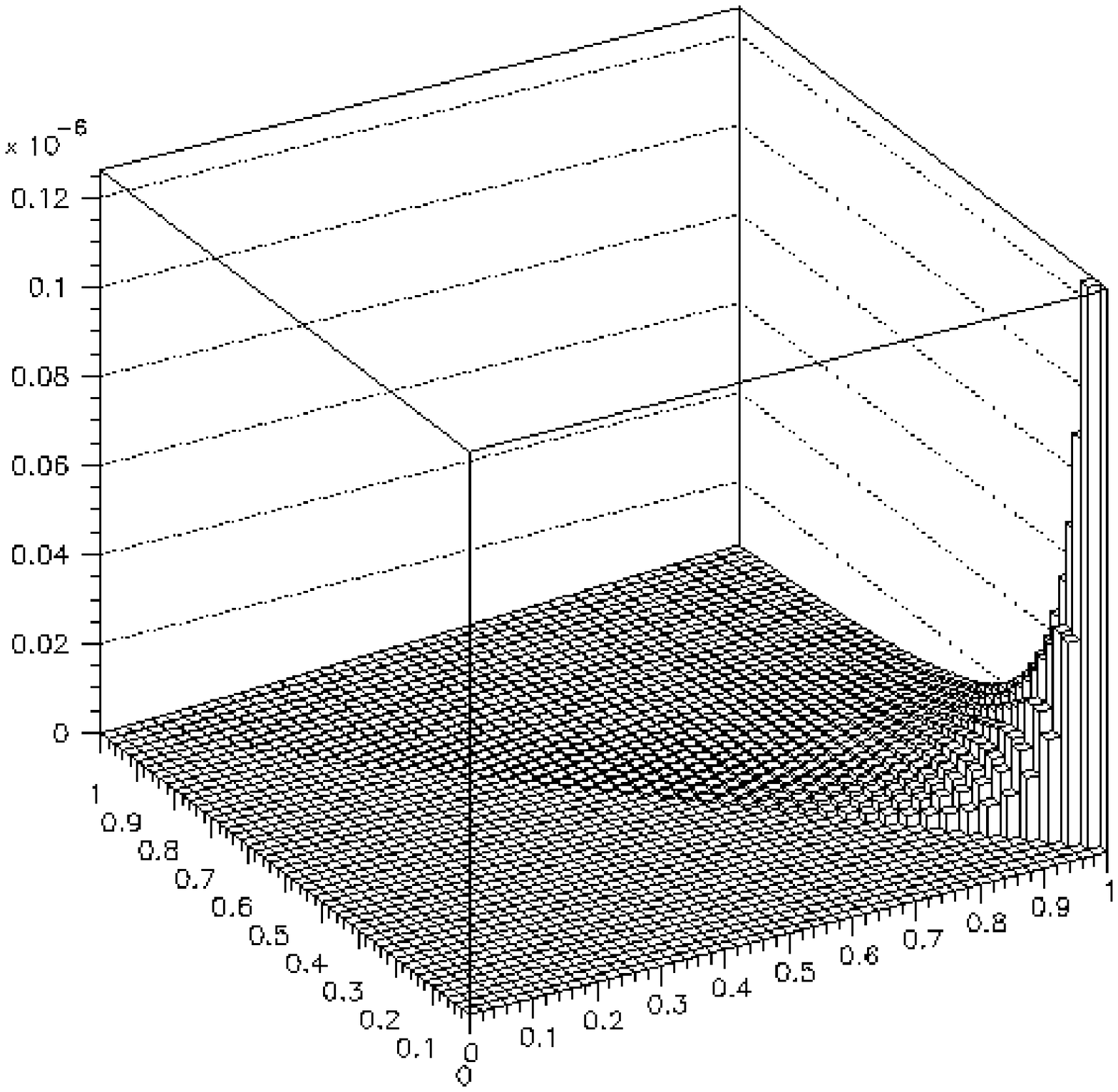,angle=0,height=8cm,width=\linewidth}
\vskip-2.0cm
\centerline{
$x_2$
\qquad\qquad\qquad\qquad
$x_1$}
\end{minipage}\hfil
\begin{minipage}[b]{.33333\linewidth}
\centerline{$+F_3$}
\vspace{-0.75truecm}\centering\epsfig{file=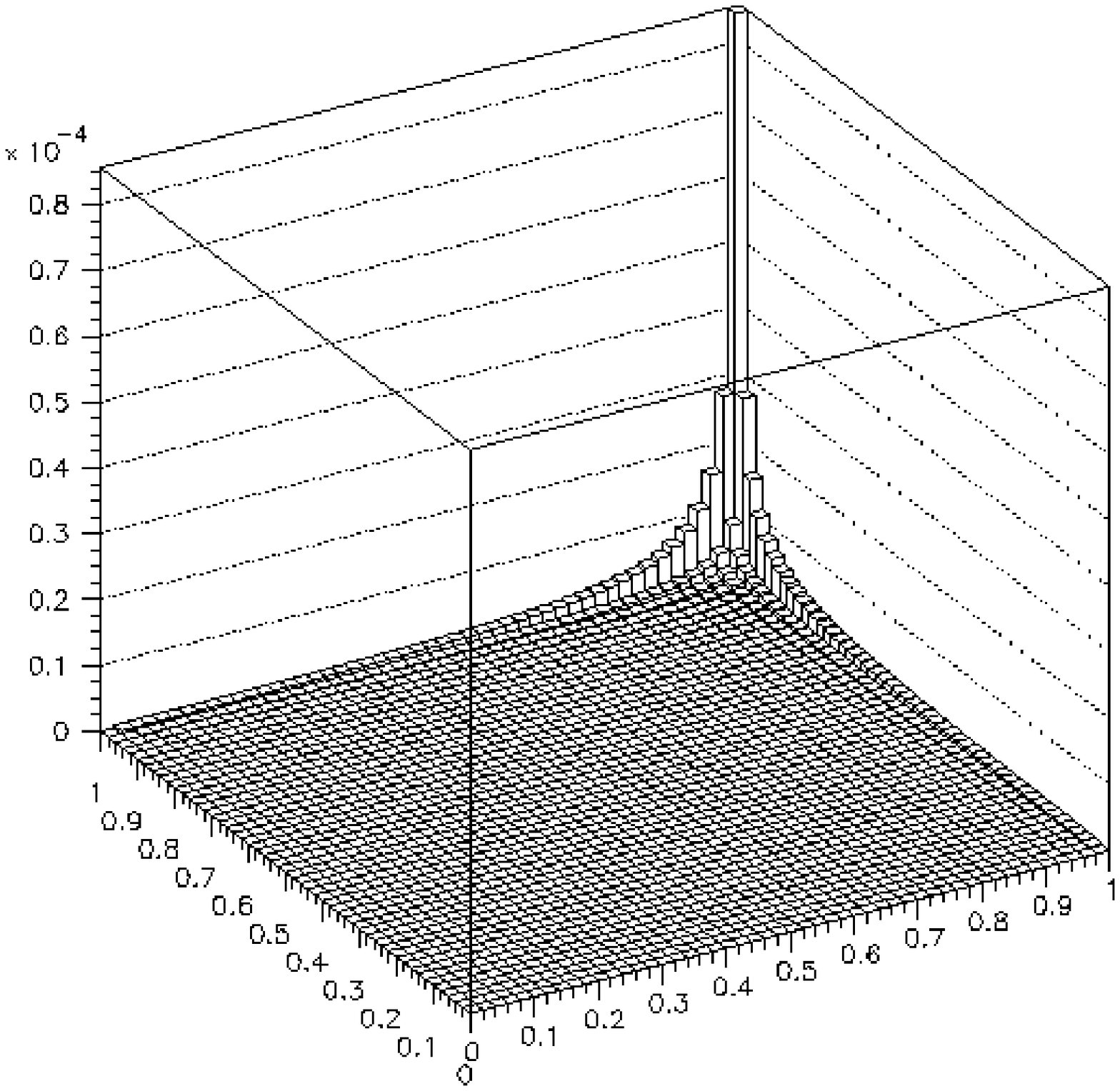,angle=0,height=8cm,width=\linewidth}
\vskip-2.0cm
\centerline{
$x_2$
\qquad\qquad\qquad\qquad
$x_1$}
\end{minipage}\hfil
\centerline{}
\begin{minipage}[b]{.33333\linewidth}
\centerline{$+F_4$}
\vspace{-0.75truecm}\centering\epsfig{file=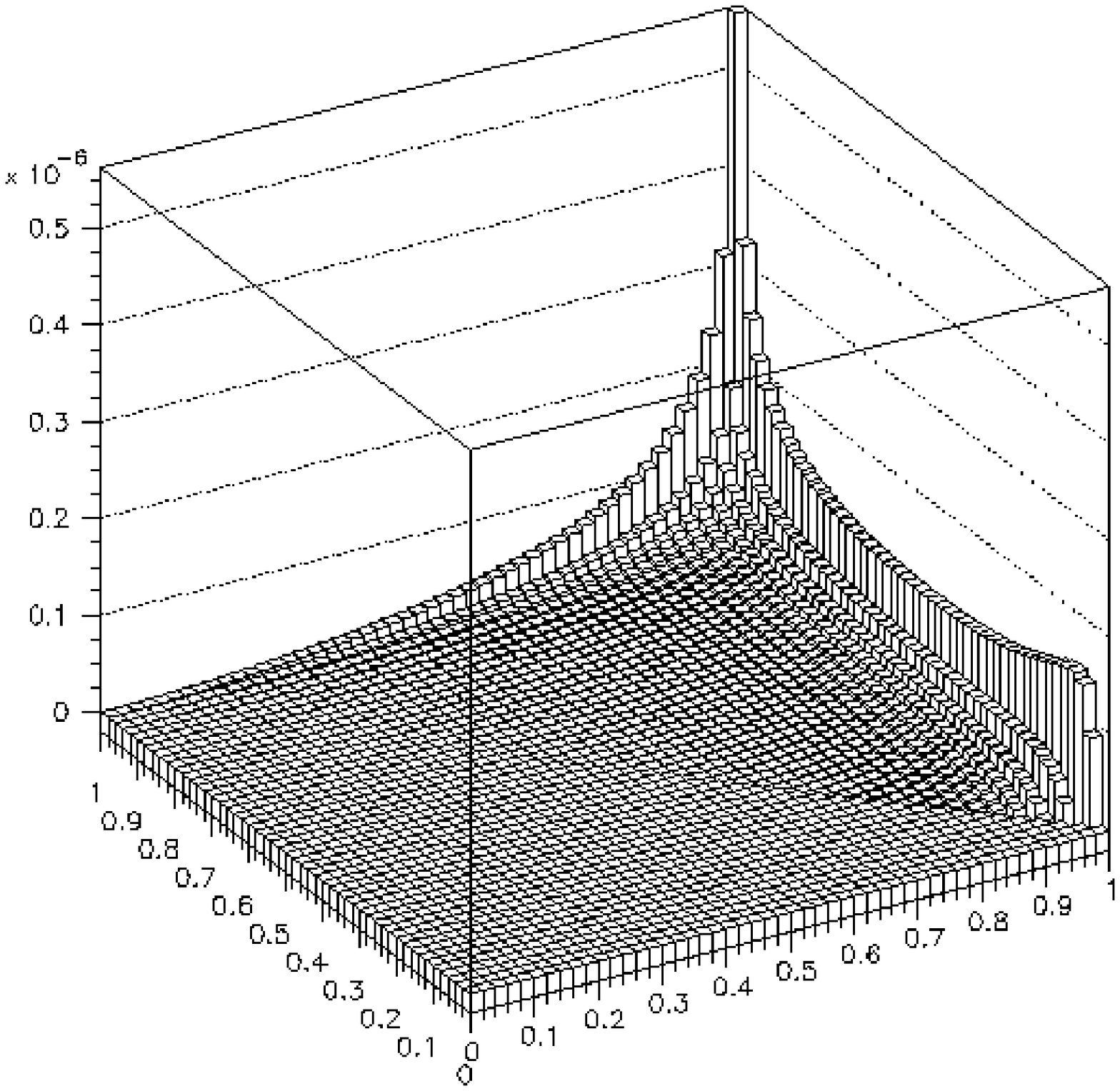,angle=0,height=8cm,width=\linewidth}
\vskip-2.0cm
\centerline{
$x_2$
\qquad\qquad\qquad\qquad
$x_1$}
\end{minipage}\hfil
\begin{minipage}[b]{.33333\linewidth}
\centerline{$-F_5$}
\vspace{-0.75truecm}\centering\epsfig{file=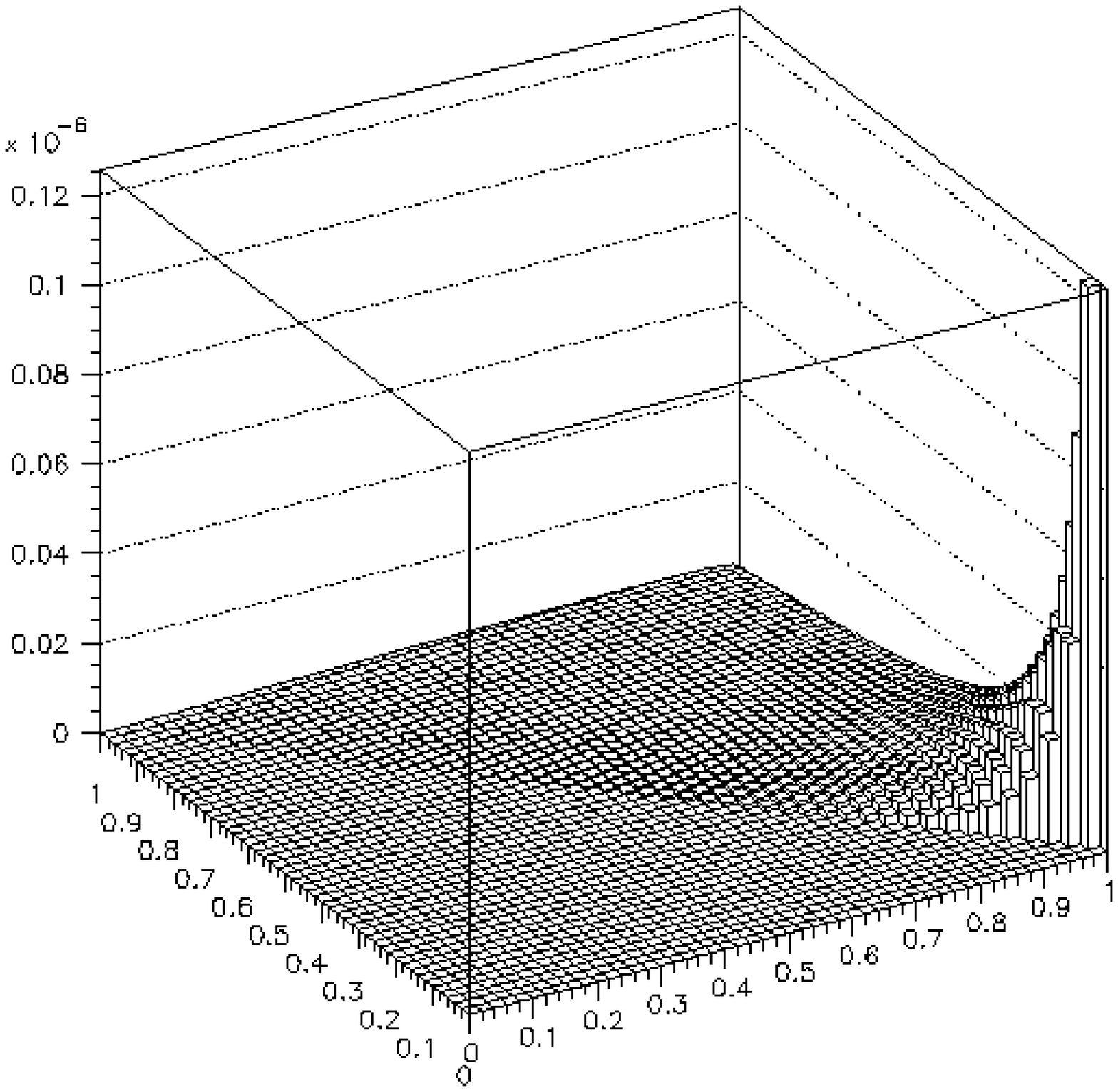,angle=0,height=8cm,width=\linewidth}
\vskip-2.0cm
\centerline{
$x_2$
\qquad\qquad\qquad\qquad
$x_1$}
\end{minipage}\hfil
\begin{minipage}[b]{.33333\linewidth}
\centerline{$-F_6$}
\vspace{-0.75truecm}\centering\epsfig{file=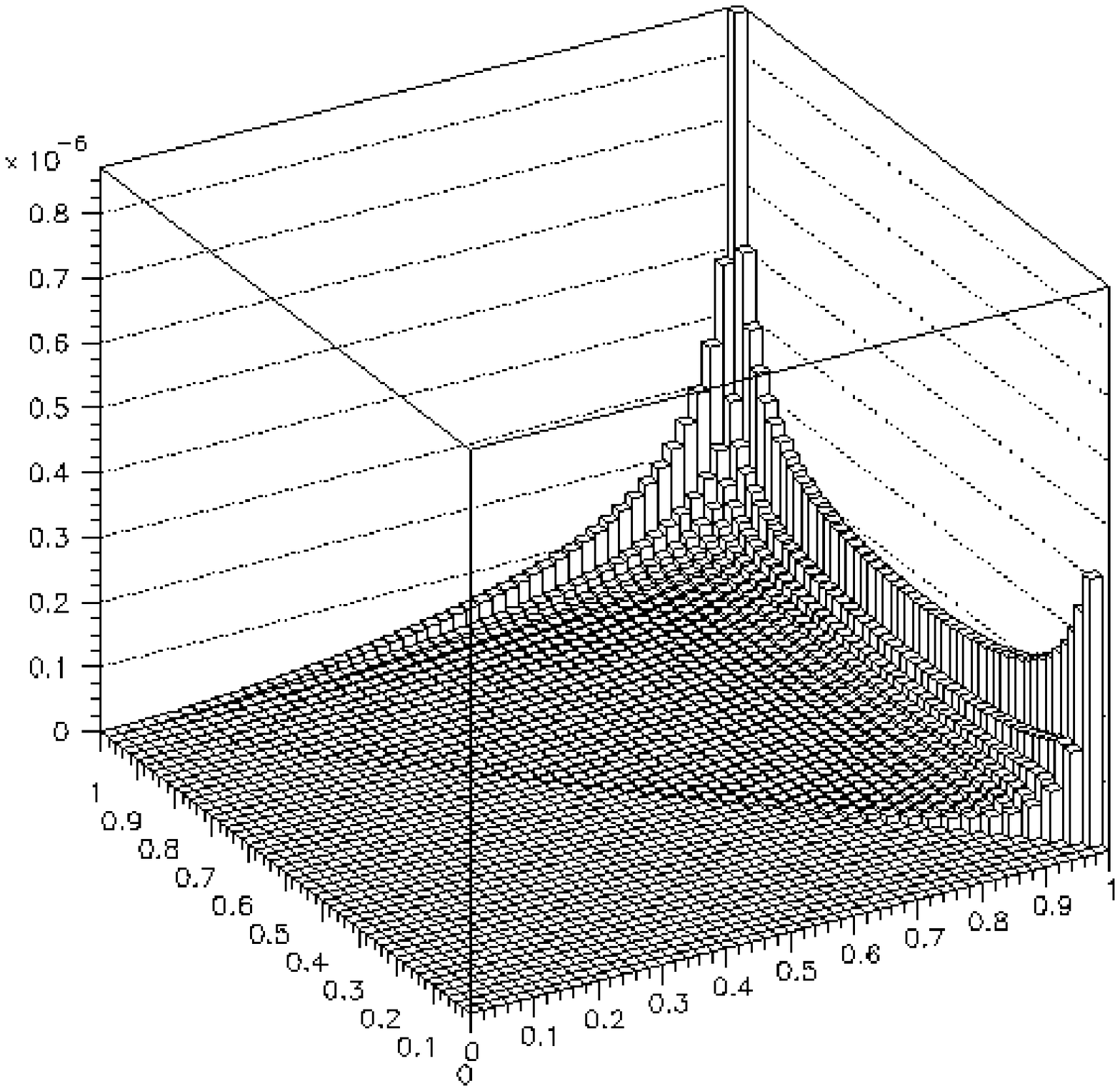,angle=0,height=8cm,width=\linewidth}
\vskip-2.0cm
\centerline{
$x_2$
\qquad\qquad\qquad\qquad
$x_1$}
\end{minipage}\hfil
\centerline{}
\begin{minipage}[b]{.33333\linewidth}
\centerline{$-F_7$}
\vspace{-0.75truecm}\centering\epsfig{file=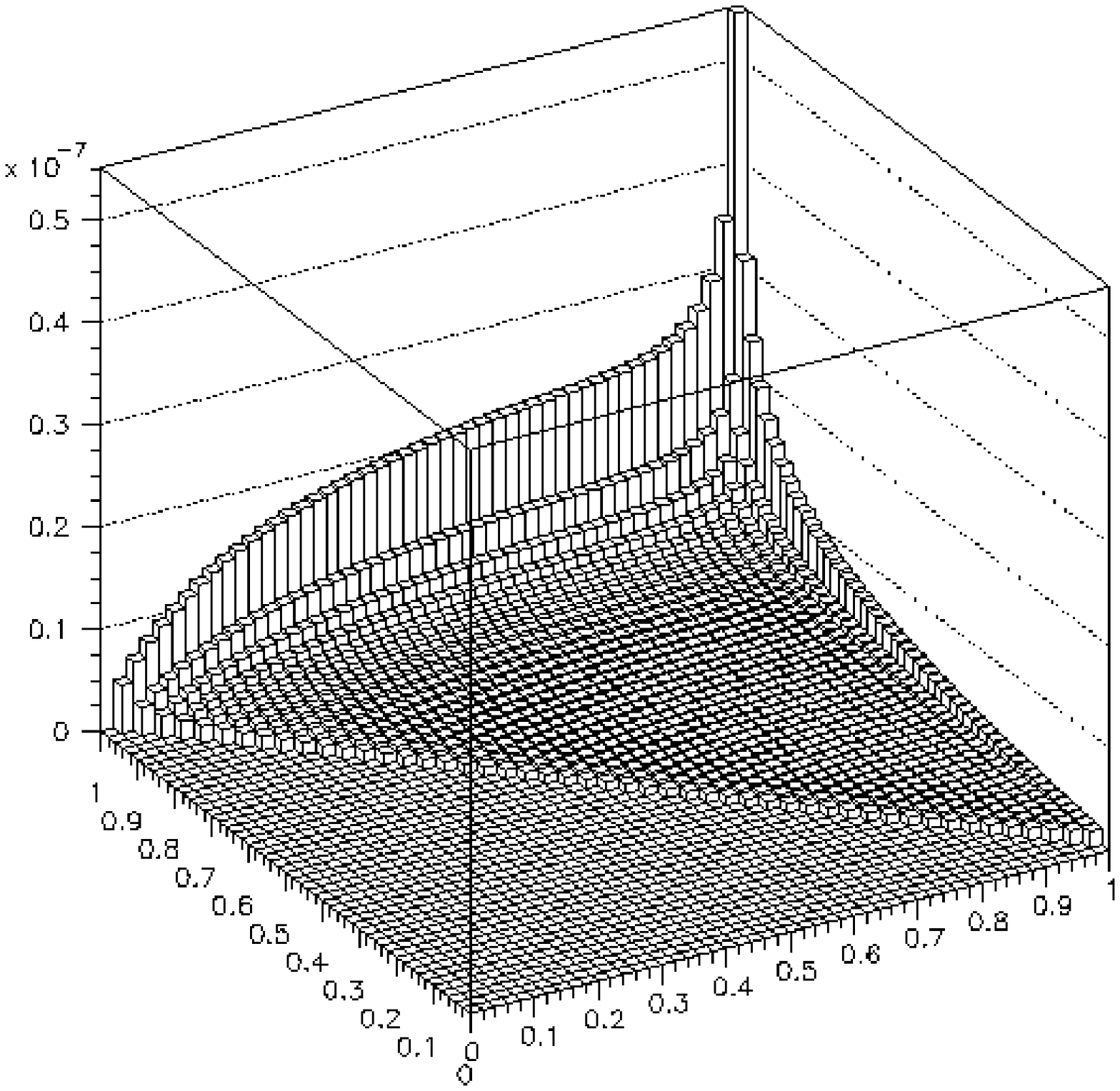,angle=0,height=8cm,width=\linewidth}
\vskip-2.0cm
\centerline{
$x_2$
\qquad\qquad\qquad\qquad
$x_1$}
\end{minipage}\hfil
\begin{minipage}[b]{.33333\linewidth}
\centerline{$-F_8$}
\vspace{-0.75truecm}\centering\epsfig{file=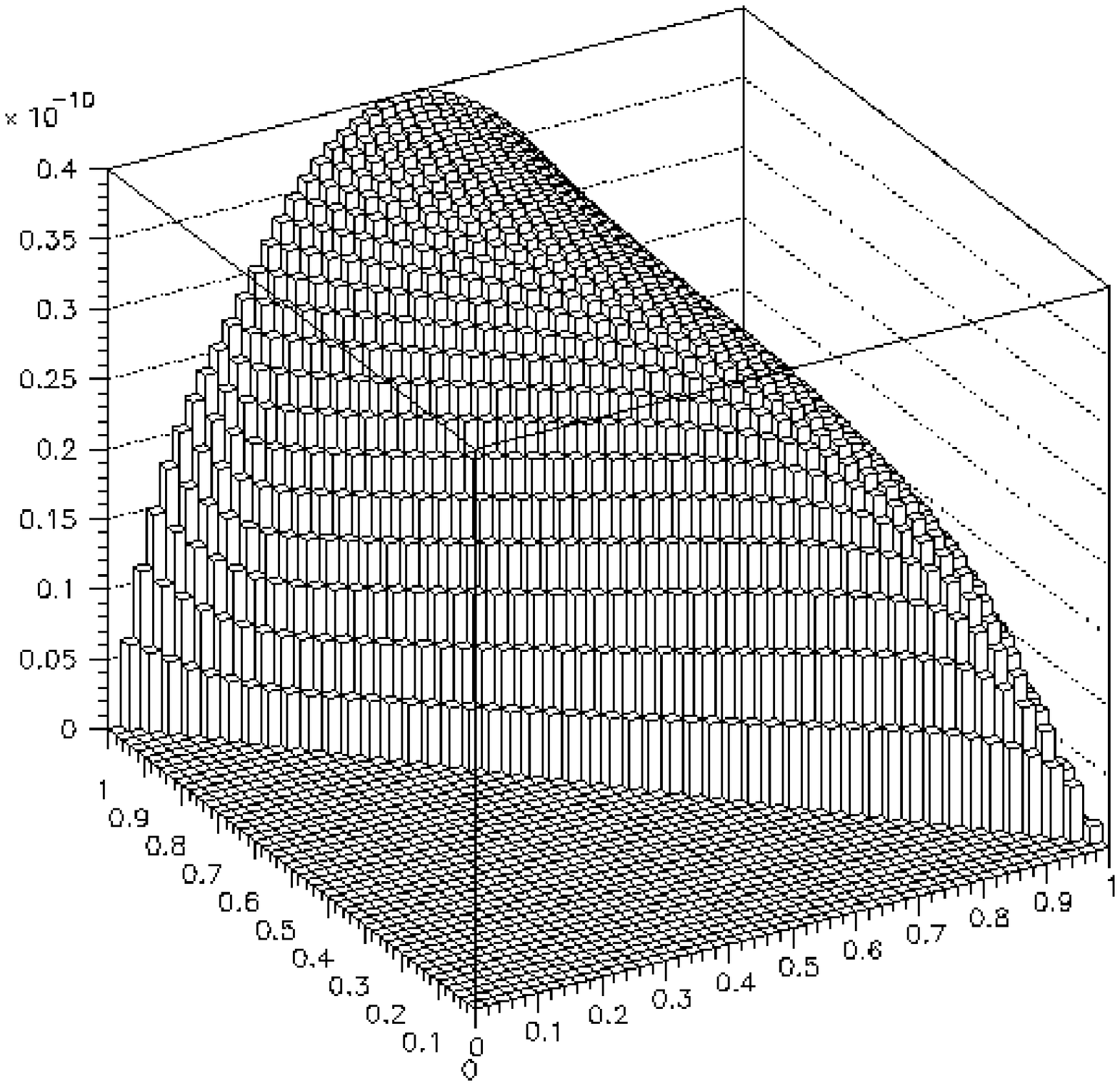,angle=0,height=8cm,width=\linewidth}
\vskip-2.0cm
\centerline{
$x_2$
\qquad\qquad\qquad\qquad
$x_1$}
\end{minipage}\hfil
\begin{minipage}[b]{.33333\linewidth}
\centerline{$-F_9$}
\vspace{-0.75truecm}\centering\epsfig{file=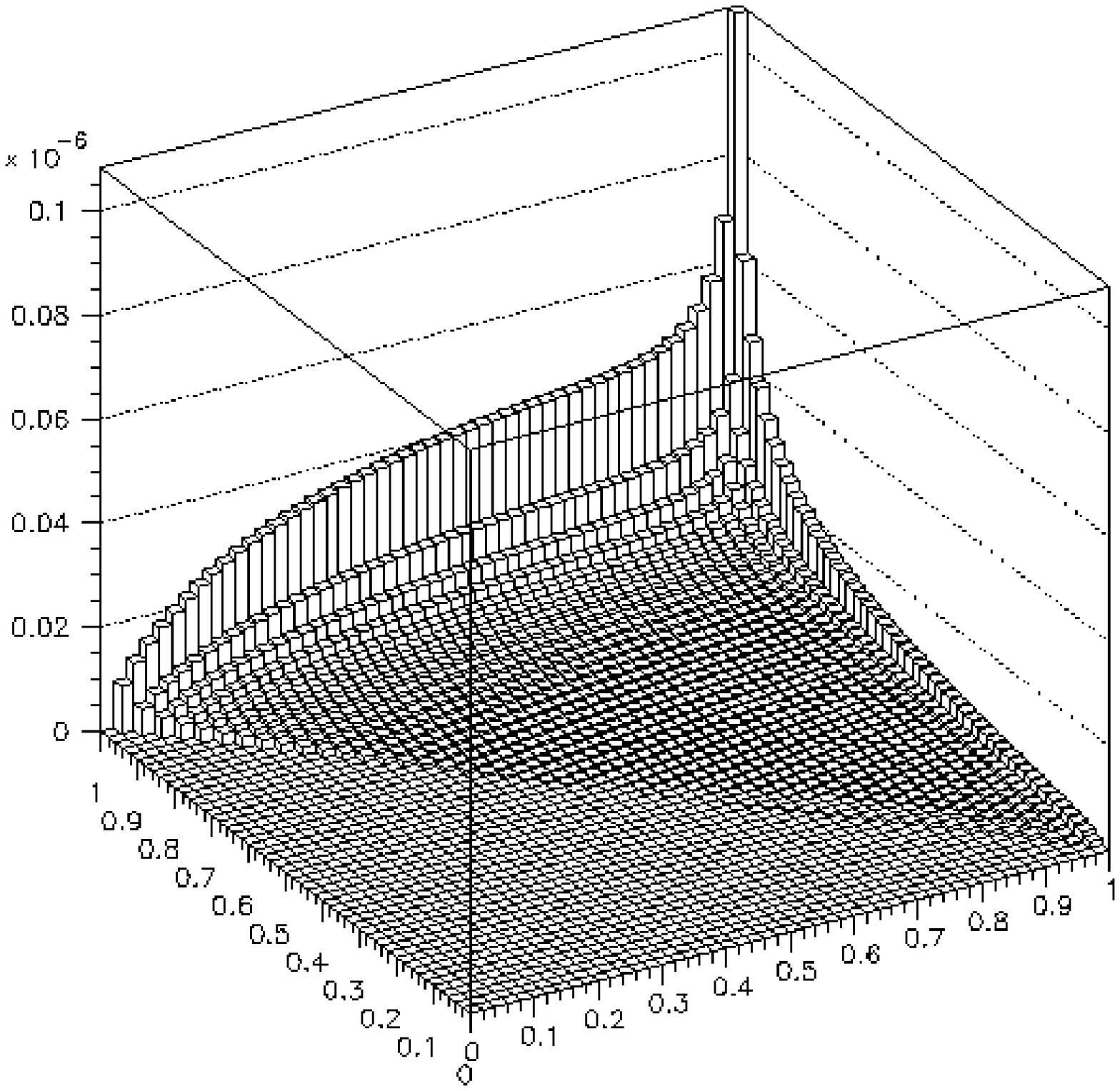,angle=0,height=8cm,width=\linewidth}
\vskip-2.0cm
\centerline{
$x_2$
\qquad\qquad\qquad\qquad
$x_1$}
\end{minipage}\hfil
\centerline{}

\noindent
\caption{\small The nine form-factors defined in eq.~(\ref{FFs}) as a function
of the antiquark ($i=1$) and quark ($i=2$) 
energy fractions $x_i=\frac{2E_i}{\sqrt s}$ at $\sqrt s=M_Z$ 
in units of nb, for left-handed incoming electrons.
(Note that in some cases we plot the opposite of the form-factor.)}
\label{fig:FFs-L}
\end{center}
\end{figure}

\begin{figure}
\begin{center}
\centerline{}
\begin{minipage}[b]{.33333\linewidth}
\centerline{$-F_1$}
\vspace{-0.75truecm}\centering\epsfig{file=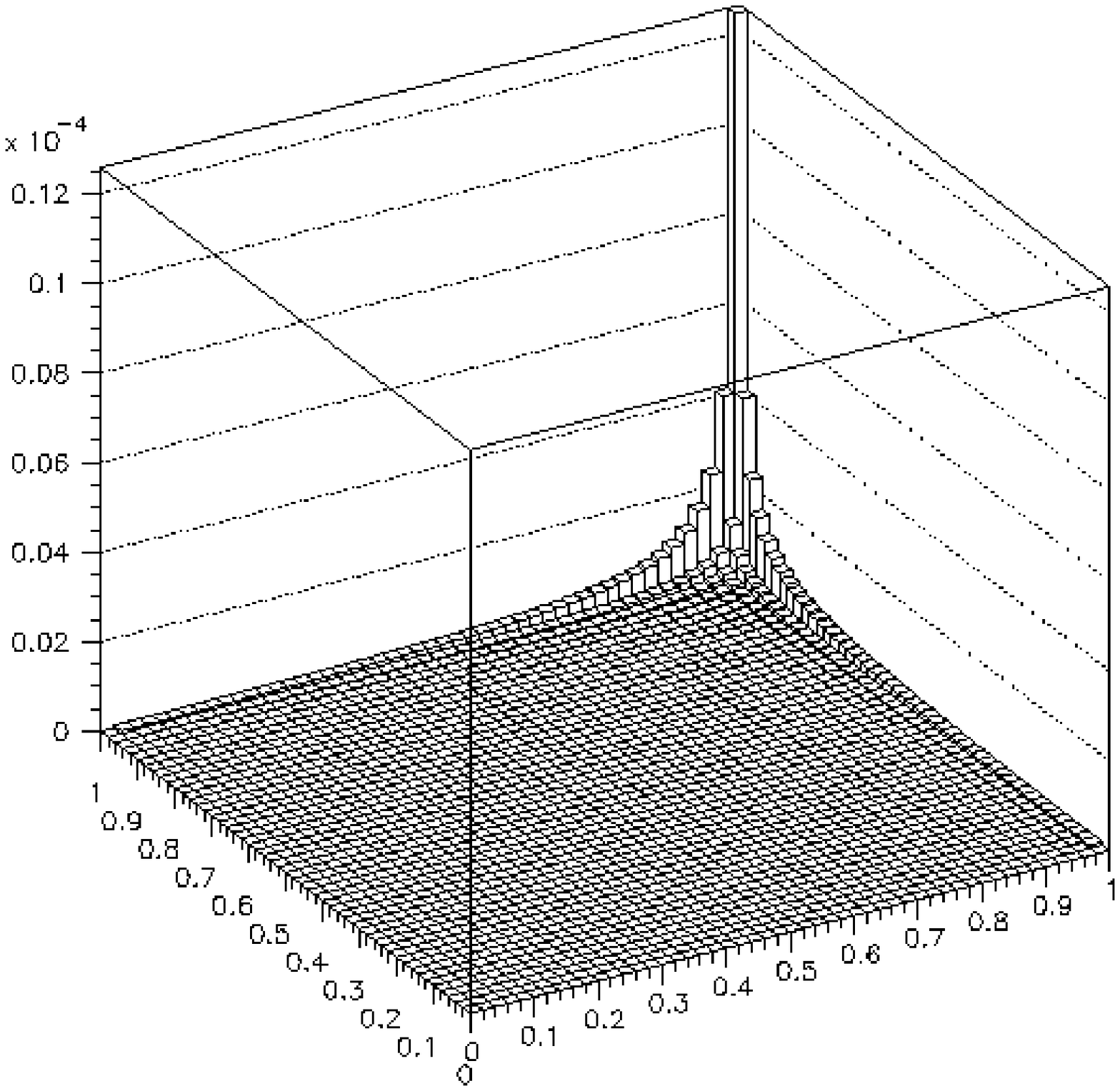,angle=0,height=8cm,width=\linewidth}
\vskip-2.0cm
\centerline{
$x_2$
\qquad\qquad\qquad\qquad
$x_1$}
\end{minipage}\hfil
\begin{minipage}[b]{.33333\linewidth}
\centerline{$-F_2$}
\vspace{-0.75truecm}\centering\epsfig{file=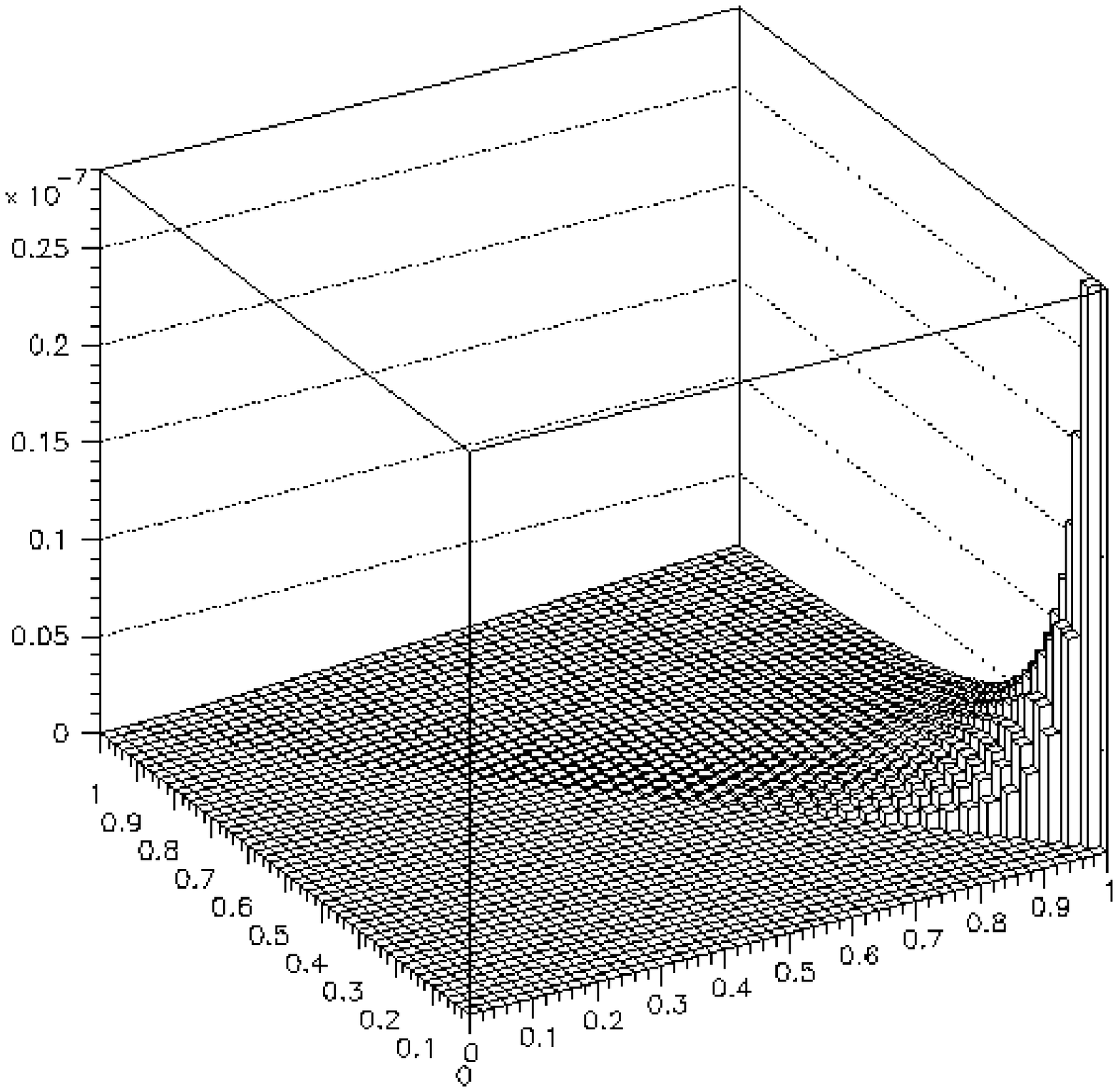,angle=0,height=8cm,width=\linewidth}
\vskip-2.0cm
\centerline{
$x_2$
\qquad\qquad\qquad\qquad
$x_1$}
\end{minipage}\hfil
\begin{minipage}[b]{.33333\linewidth}
\centerline{$-F_3$}
\vspace{-0.75truecm}\centering\epsfig{file=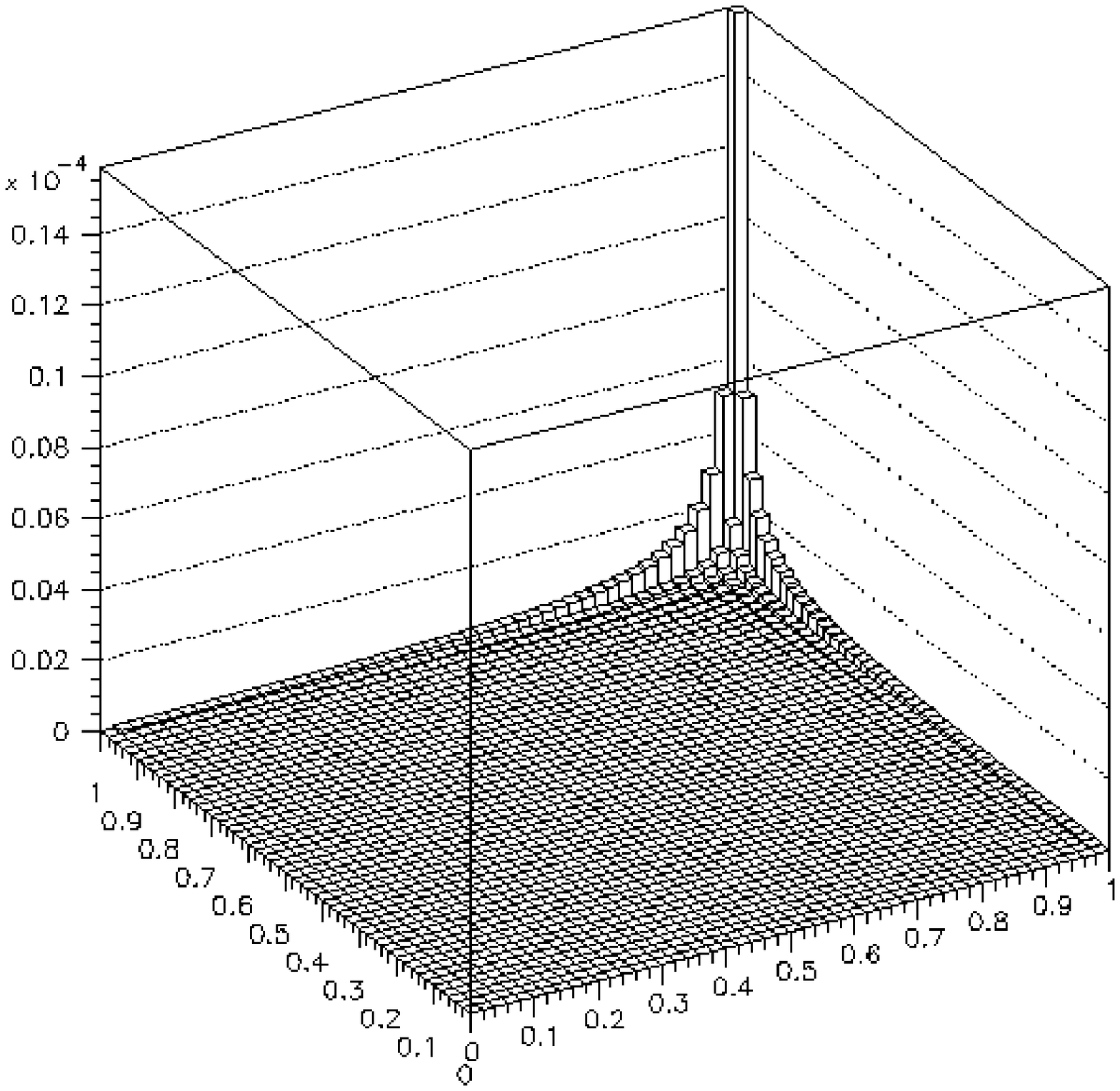,angle=0,height=8cm,width=\linewidth}
\vskip-2.0cm
\centerline{
$x_2$
\qquad\qquad\qquad\qquad
$x_1$}
\end{minipage}\hfil
\centerline{}
\begin{minipage}[b]{.33333\linewidth}
\centerline{$+F_4$}
\vspace{-0.75truecm}\centering\epsfig{file=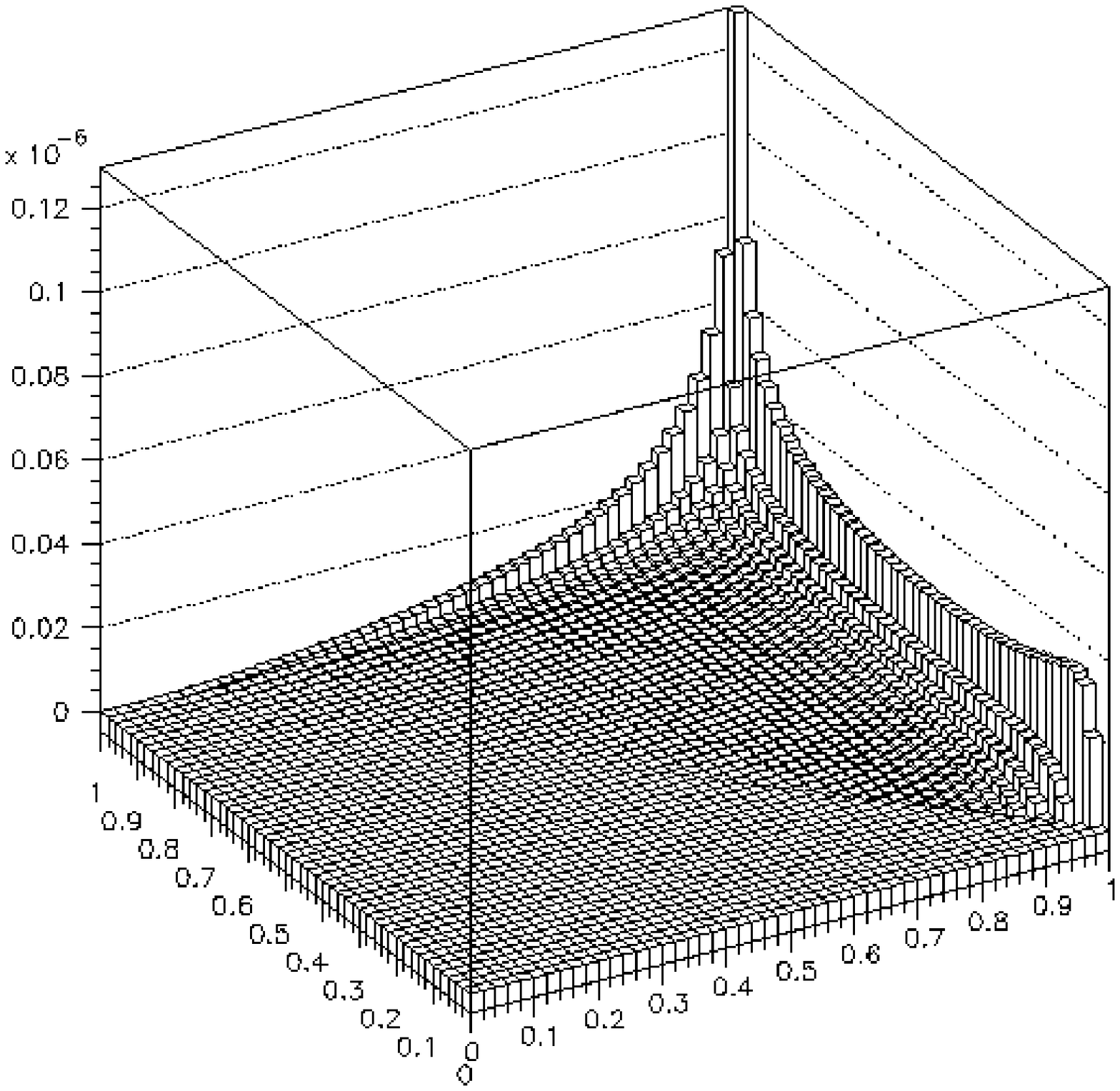,angle=0,height=8cm,width=\linewidth}
\vskip-2.0cm
\centerline{
$x_2$
\qquad\qquad\qquad\qquad
$x_1$}
\end{minipage}\hfil
\begin{minipage}[b]{.33333\linewidth}
\centerline{$+F_5$}
\vspace{-0.75truecm}\centering\epsfig{file=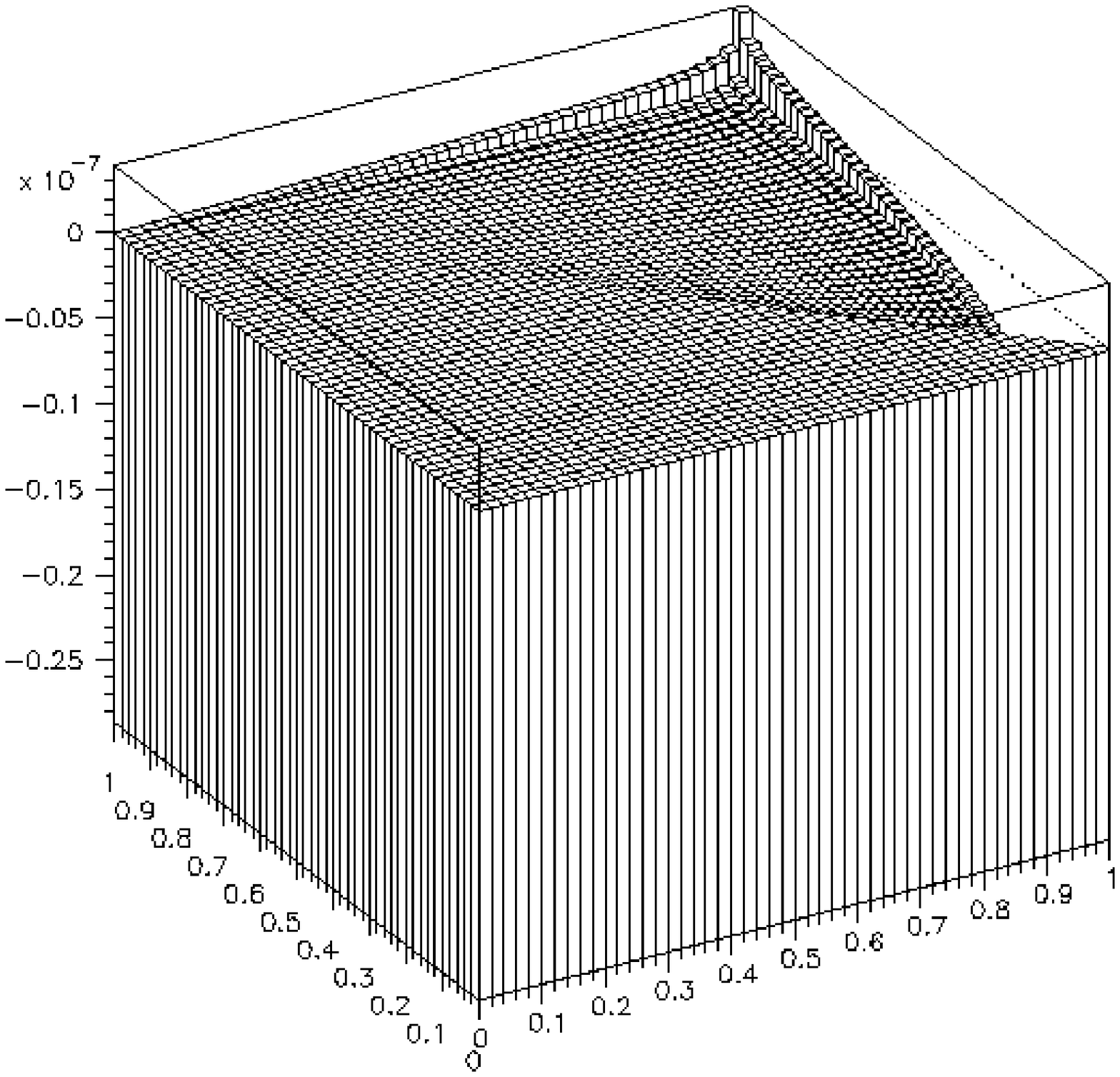,angle=0,height=8cm,width=\linewidth}
\vskip-2.0cm
\centerline{
$x_2$
\qquad\qquad\qquad\qquad
$x_1$}
\end{minipage}\hfil
\begin{minipage}[b]{.33333\linewidth}
\centerline{$+F_6$}
\vspace{-0.75truecm}\centering\epsfig{file=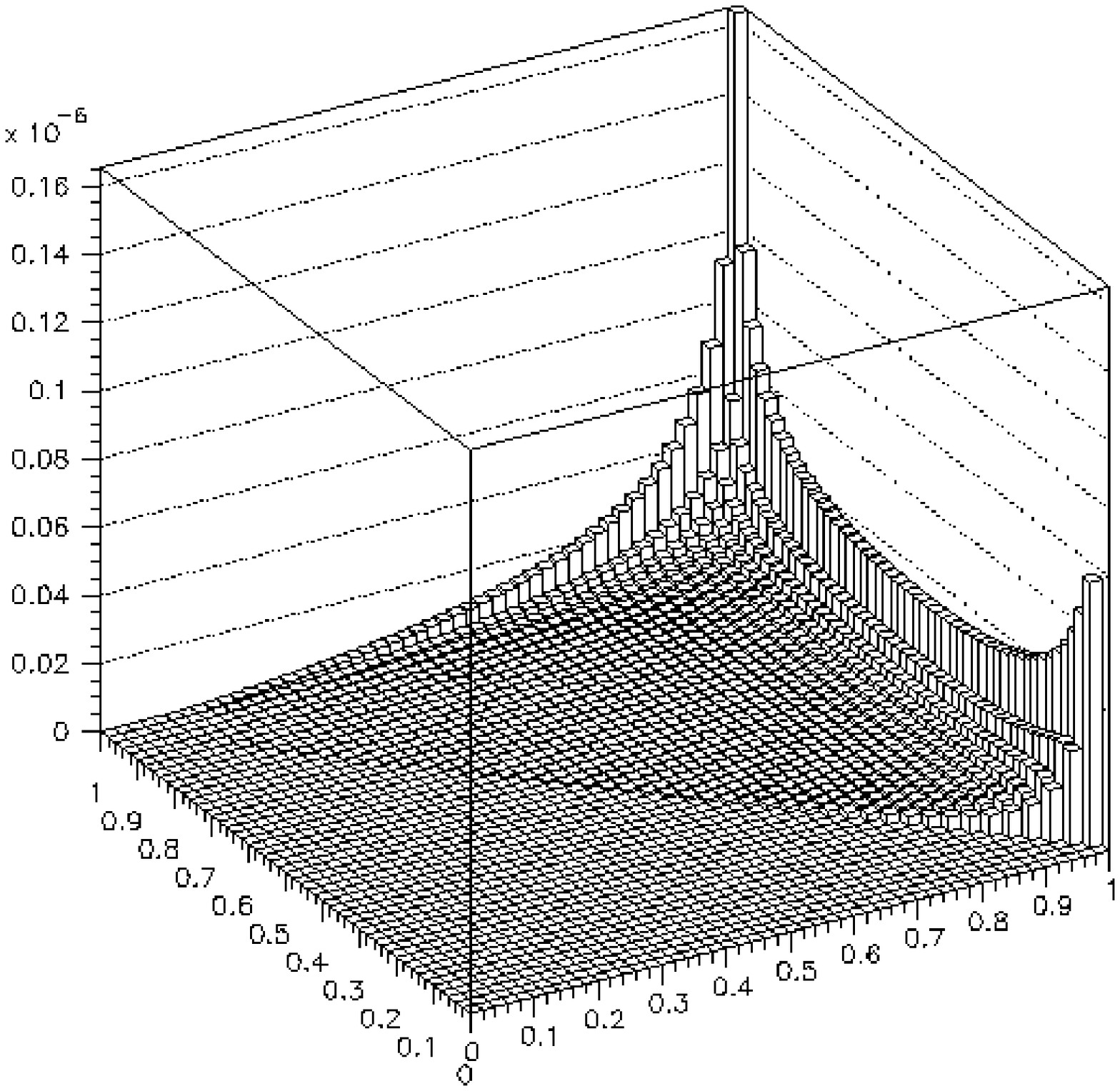,angle=0,height=8cm,width=\linewidth}
\vskip-2.0cm
\centerline{
$x_2$
\qquad\qquad\qquad\qquad
$x_1$}
\end{minipage}\hfil
\centerline{}
\begin{minipage}[b]{.33333\linewidth}
\centerline{$-F_7$}
\vspace{-0.75truecm}\centering\epsfig{file=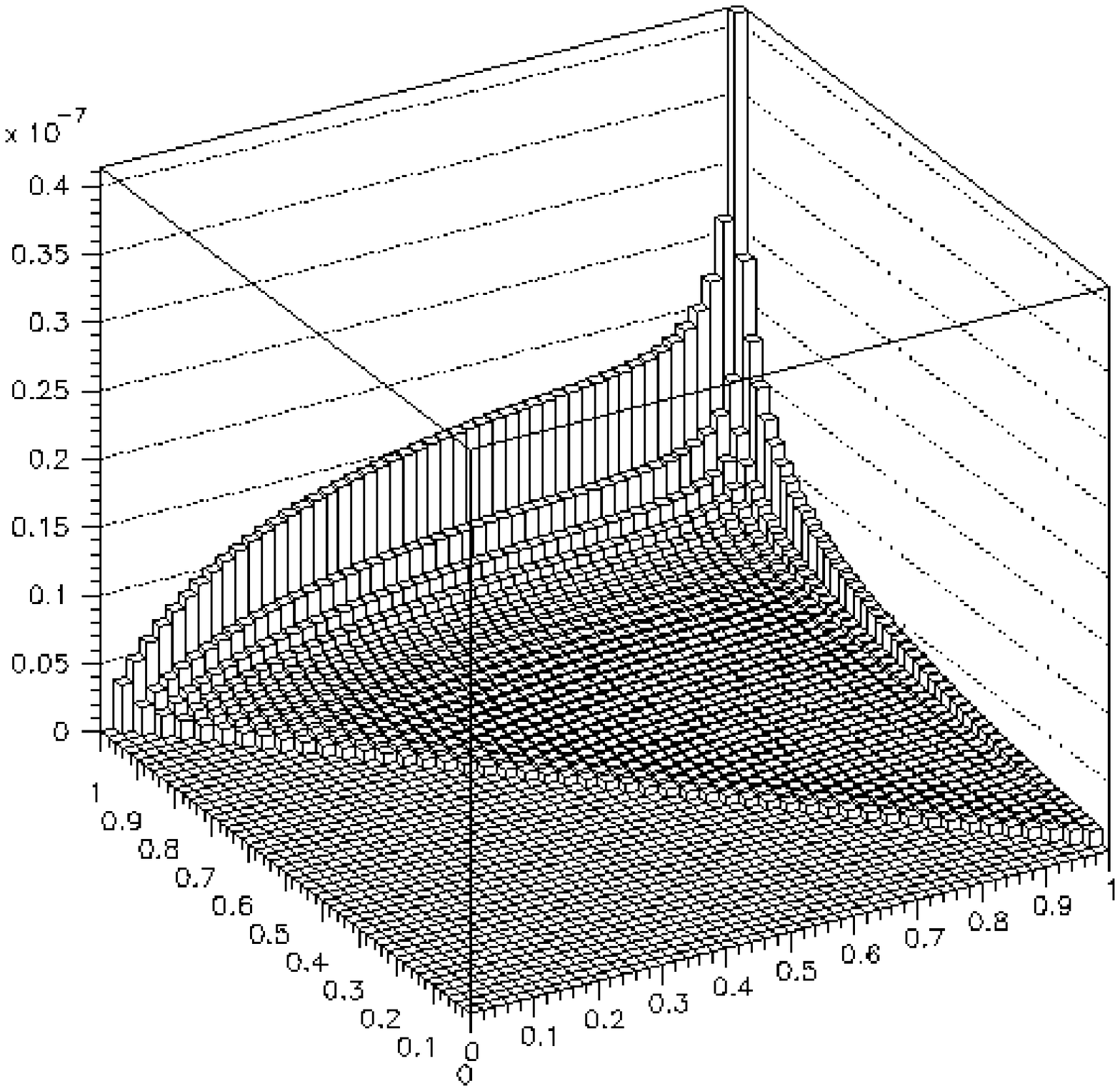,angle=0,height=8cm,width=\linewidth}
\vskip-2.0cm
\centerline{
$x_2$
\qquad\qquad\qquad\qquad
$x_1$}
\end{minipage}\hfil
\begin{minipage}[b]{.33333\linewidth}
\centerline{$-F_8$}
\vspace{-0.75truecm}\centering\epsfig{file=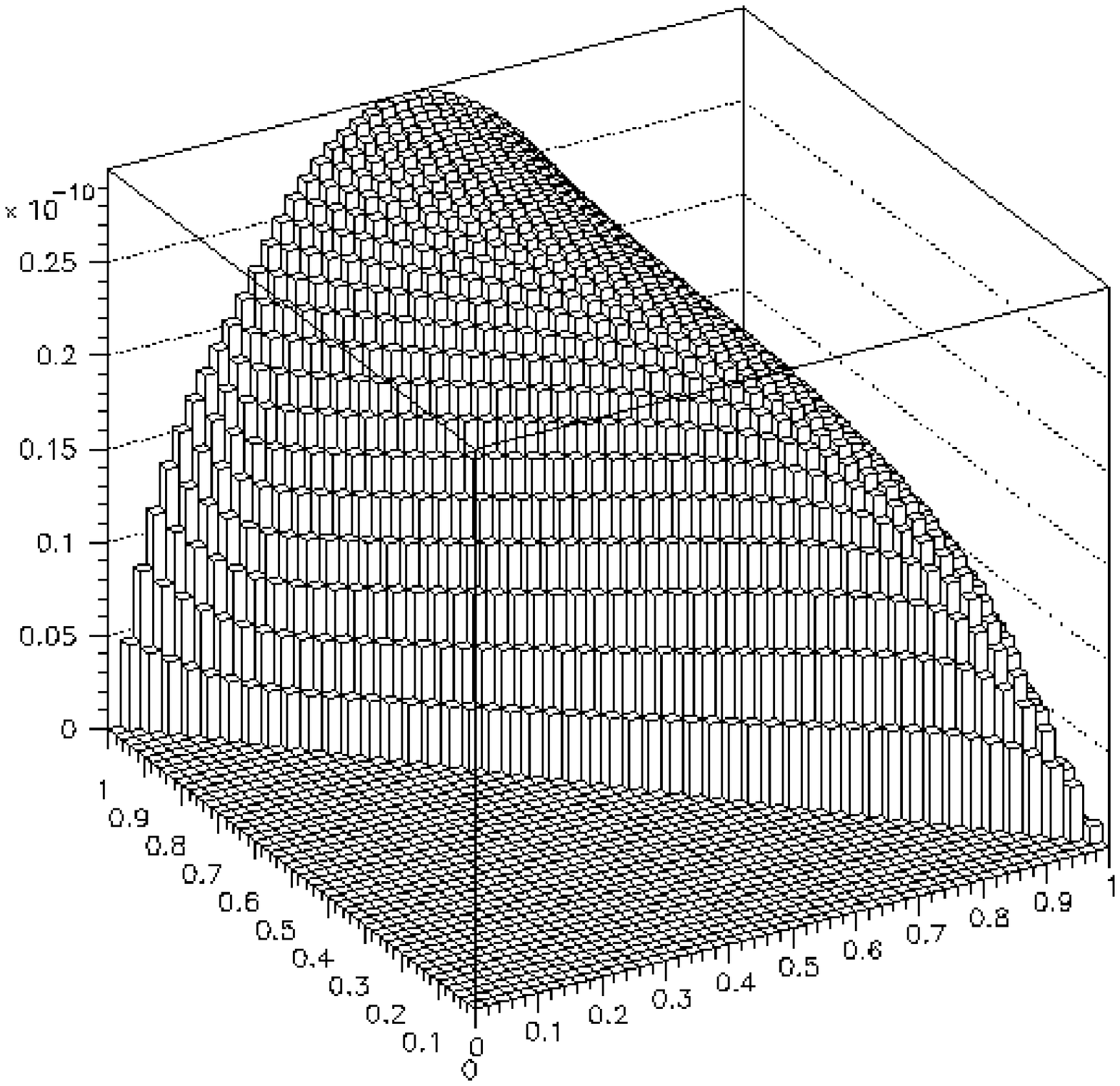,angle=0,height=8cm,width=\linewidth}
\vskip-2.0cm
\centerline{
$x_2$
\qquad\qquad\qquad\qquad
$x_1$}
\end{minipage}\hfil
\begin{minipage}[b]{.33333\linewidth}
\centerline{$+F_9$}
\vspace{-0.75truecm}\centering\epsfig{file=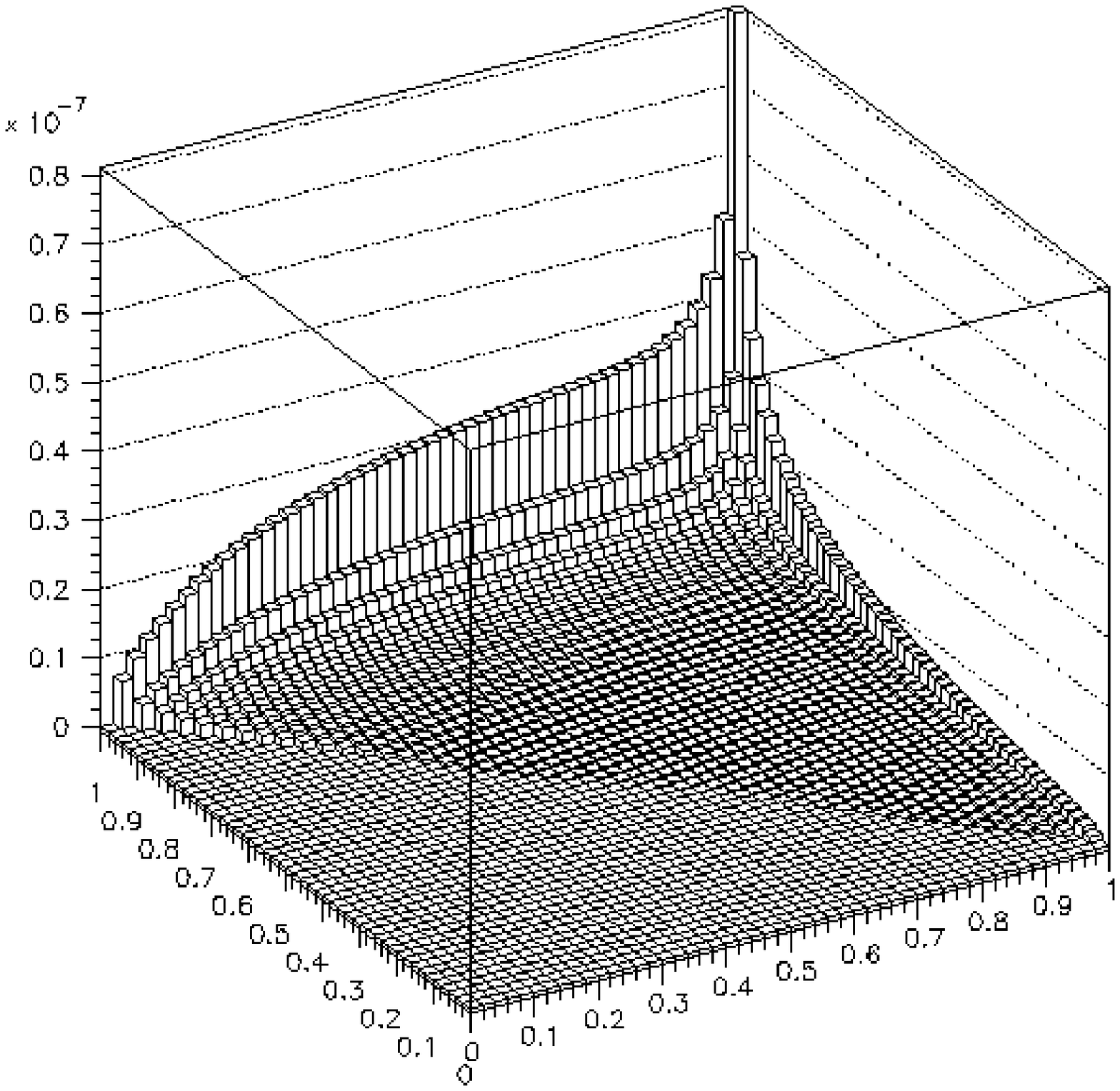,angle=0,height=8cm,width=\linewidth}
\vskip-2.0cm
\centerline{
$x_2$
\qquad\qquad\qquad\qquad
$x_1$}
\end{minipage}\hfil
\centerline{}

\noindent
\caption{\small The nine form-factors defined in eq.~(\ref{FFs}) as a function
of the antiquark ($i=1$) and quark ($i=2$) 
energy fractions $x_i=\frac{2E_i}{\sqrt s}$ at $\sqrt s=M_Z$ 
in units of nb, for right-handed incoming electrons.
(Note that in some cases we plot the opposite of the form-factor.)}
\label{fig:FFs-R}
\end{center}
\end{figure}

The ability to polarise electron (and possibly, positron) beams joined
with the high luminosity available
render future LCs a privileged environment in which to test the
structure of hadronic samples. As noted earlier,
differential spectra may well carry the distinctive hallmark of some new
and heavy strongly interactive particles (such as squarks
and gluinos in Supersymmetry), whose rest mass is too
large for these to be produced in pairs
as real states but that may enter as virtual 
objects into multi-jet events. 
Similar effects may however also be induced by the NLO-W corrections tackled 
here. 
Observables where such effects would immediately be evident are 
the so-called the `unintegrated' (or `oriented') Thrust 
distributions associated to each of the form-factors in
eq.~(\ref{FFs}) (wherein $S={\rm T}$), as can be seen 
in Fig.~7 of Ref.~\cite{Maina:2002wz}. Here, 
as a benchmark for future studies along the above lines, we directly
reproduce in Figs.~\ref{fig:FFs-L}--\ref{fig:FFs-R} the differential
structure of the nine form-factors given in eq.~(\ref{FFs}),
as a function of the energy fractions $x_1$ and $x_2$, at $\sqrt
s=M_Z$ (at this energy, the shape is basically the 
same for both final states in (\ref{procj})--(\ref{procb})), 
for the case of a left- and right-handed 
incoming electrons, respectively. Here,
we should mention that the NLO-W corrections to the
corresponding non-zero tree-level distributions were found of the order
$-$(1--2)\% on average for the case of the unflavoured sample and twice
as much for the case of the $b$-quark one, with the corrections 
arising almost entirely from the left-handed incoming electrons\footnote{For 
reason of 
space, we refrain
from presenting here the LO dependence of $F_1,...F_6$ in term
of $x_1$ and $x_2$. This can easily be reproduced starting from
the formulae in \cite{Maina:2002wz}.}. Finally notice that all nine 
distributions in Figs.~\ref{fig:FFs-L}--\ref{fig:FFs-R}, 
in the presence of a precise determination
of $\alpha$ and $\beta$ (or, for that matter,
any other combination of angles used to parametrise the event orientation),
are directly observable in the case the
$\bar bbg$ final state if also the charge (other than the flavour)
of the quark is known, e.g., via high-$p_T$ lepton tagging
from $B$-meson decays or via global jet charge determination.
If not, all distributions in Figs.~\ref{fig:FFs-L}--\ref{fig:FFs-R}
have to be symmetrised around the $x_1=x_2$ direction. In the case
of the full hadronic sample $\bar qqg$, when no flavour tagging
is available, one normally identifies the two most energetic jets
in the three-jet sample with those originating from the
quark-antiquark pair. Although not always true, this approach
is known to be a good approximation for most studies (see, e.g.,
Ref.~\cite{ordering}). Hence, even in this case one may be able
to verify to a good approximation
the shape of the form-factors $F_1,...F_9$ in terms
of the energy fractions.

\section{Summary}
\label{Sec:Conclusions}

In conclusion, we have found that, at $\sqrt s=M_Z$, the size of the 
NLO-W corrections to three-jet rates and event shapes is
rather small, of order percent or so, hence confirming that determinations
of $\as$ at LEP1 and SLC are stable in this respect and that the
SM background to parity-violating effects possibly induced by new physics
is well under control. In contrast, NLO-W effects 
ought to be included in the case
of future high-luminosity LCs running at the $Z$ pole, such as GigaZ,
where the accuracy of $\as$  measurements from such observables
is expected to reach the
$0.1\%$ level. Effects from NLO-W corrections are somewhat larger in the case
of $b$-quarks in the final state, in comparison to the case in which
all flavours are included in the hadronic sample, because of the
presence of the top quark in the one-loop virtual contributions. 

Since the exploitation of beam polarisation effects will be a
key feature of experimental analyses of hadronic events at future LCs,
we have computed the full differential structure of three-jet processes
in the presence of polarised electrons and positrons, in terms of the energy 
fractions of the two leading jets and of two angles describing the final
state orientation. 
The cross-sections were then parametrised by means of nine independent 
form-factors, the latter presented as a function of 
the (anti)quark energy fractions. 
Three of these form-factors carry parity-violating
effects which cannot then receive contributions from ordinary QCD.
For the two that are non-zero at LO, i.e., $F_3$ and $ F_6$,
the NLO-W 
corrections were found as large as $4\%$, for
the case of $b$-quarks. Such higher-order weak effects
should appropriately be subtracted from hadronic samples in the search for 
physics beyond the SM.

\section*{Acknowledgements} 
SM thanks the conveners
of the `Top and QCD' working group for their kind invitation.

\newpage

\newpage

\newpage

\newpage

\newpage

\newpage

\newpage

{\small
\begin{thebibliography}{99}

\bibitem{LCs}
K.~Abe {\it et al.}, [The ACFA Linear Collider Working Group],
{\tt hep-ph/0109166};
T.~Abe {\it et al.}, [The American Linear Collider Working Group],
{\tt hep-ex/0106055}; {\tt hep-ex/0106056}; {\tt hep-ex/0106057};
{\tt hep-ex/0106058};
J.A. Aguilar-Saavedra {\it et al.}, [The 
ECFA/DESY LC Physics Working Group],  {\tt hep-ph/0106315};
G. Guignard (editor), [The CLIC Study Team], preprint CERN-2000-008 (2000).

\bibitem{Winter} M. Winter, LC Note LC-PHSM-2001-016, February 2001
(and references therein).

\bibitem{Maina:2002wz}
E.~Maina, S.~Moretti and D.~A.~Ross,
JHEP {\bf 04} (2003) 056.

\bibitem{BDS} A. Brandenburg, L. Dixon and Y. Shadmi,
\prd{53} {1996} {1264}.

\bibitem{KS} J.G. K\"{o}rner and G. Schuler, \zpc {26} {1985} {559};
~K. Hagiwara, T. Kuruma and Y. Yamada,
{Nucl. Phys.} {\bf B358} (1991) 80.

\bibitem{BMM}A.~Ballestrero, E.~Maina and S.~Moretti,
{Phys.\ Lett.}  {\bf B294} (1992) 425;
{Nucl. Phys.} {\bf B415} (1994) 265.

\bibitem{bbgNLO}
G.~Rodrigo, A.~Santamaria and M.~Bilenky, 
{Phys. Rev. Lett.} {\bf 79} (1997) 193;
J. Phys. {\bf G25} (1999) 1593;
{\tt hep-ph/9802359};\\
 G.~Rodrigo, {{\tt hep-ph/9703359}}; {Nucl. Phys. Proc. Suppl.}
  {\bf 54A} (1997) 60;\\
W.~Bernreuther, A.~Brandenburg and P.~Uwer, 
{Phys. Rev. Lett.} {\bf 79} (1997) 189;\\
A.~Brandenburg and P.~Uwer, {Nucl. Phys.} {\bf B515} (1998) 279;\\
P.~Nason and C.~Oleari, {Phys. Lett.} {\bf B407} (1997) 57;
{Nucl. Phys.} {\bf B521} (1998) 237.

\bibitem{ERT} R.K. Ellis, D.A. Ross and A.E. Terrano, 
{Nucl. Phys.} {\bf B178} ({1981}) {421}.

\bibitem{schemes} S. Moretti, L. L\"onnblad and T. Sj\"ostrand,
\jhep {08} {1998} {001}.

\bibitem{jade}
JADE Collaboration, \zpc {33} {1986} {23};\\ 
S.~Bethke, {Habilitation thesis}, preprint LBL 50-208 (1987).

\bibitem{durham}
Yu.L.\ Dokshitzer, contribution cited in the `Report of the 
Hard QCD Working Group', in
Proceedings of the workshop `Jet Studies at LEP and HERA',
       Durham, December 1990, {J. Phys} {\bf G17} (1991) 1537;\\
S.~Catani, Yu.L.~Dokshitzer, M.~Olsson, G.~Turnock and B.R.~Webber,
\plb  {269} {1991} {432}.

\bibitem{cambridge} Yu.L.~Dokshitzer, G.D.~Leder, S.~Moretti and
  B.R.~Webber, \jhep {08} {1997} {001}.

\bibitem{BKSS}
S.~Bethke, Z.~Kunszt, D.E.~Soper and W.J.~Stirling,
{Nucl. Phys.}
 {\bf B370} ({1992}) {310}; Erratum, {\tt hep-ph/{9803267}}.

\bibitem{EERAD} F.A. Berends, W.T. Giele and H. Kuijf,
{Nucl. Phys.} {\bf B321} (1989) 595; W.T. Giele 
and E.W.N. Glover, \prd {46} {1992} {1980}.

\bibitem{thrust} E. Fahri, \prl {39} {1977} {1587}.

\bibitem{ordering} See, e.g.: T. Hebbeker,
{Phys. Rep.} {\bf 217} (1992) 69 (and references therein).


\end{thebibliography}
}
\end{document}